\documentclass[12pt]{iopart}

\usepackage{graphicx}
\usepackage{hyperref}
\usepackage{rotating}
\usepackage{subcaption}
\usepackage{cite}

\usepackage{color}
\usepackage{xcolor}

\usepackage{wrapfig,booktabs}

\definecolor{phlox}{rgb}{0.87, 0.0, 1.0}

\newcommand{\tum}{\ensuremath{0}}
\newcommand{\adj}{\ensuremath{{\rm adj}}}

\begin{document}

\title[Interacting Run-and-Tumble Random Walkers]{Exact Solution of Two Interacting Run-and-Tumble Random Walkers with Finite Tumble Duration}

\author{A. B. Slowman, M. R. Evans and R. A. Blythe}
\address{SUPA, School of Physics and Astronomy, University of Edinburgh, Peter Guthrie Tait Road, Edinburgh EH9 3FD}
\date{10th April 2017}
\begin{abstract}

We study a model of interacting run-and-tumble random walkers operating under mutual hardcore exclusion on a one-dimensional lattice with periodic boundary conditions. We incorporate a finite, Poisson-distributed, tumble duration so that a particle remains stationary whilst tumbling, thus generalising the persistent random walker model. We present the exact solution for the nonequilibrium stationary state of this system in the case of two random walkers. We find this to be characterised by two lengthscales, one arising from the jamming of approaching particles, and the other from one particle moving when the other is tumbling. The first of these lengthscales vanishes in a scaling limit where the continuous-space dynamics is recovered whilst the second remains finite. Thus the nonequilibrium stationary state reveals a rich structure of attractive, jammed and extended pieces.
\end{abstract}

\maketitle

\section{Introduction}

The problem of bacterial dynamics sits at the crossroads of non-equilibrium statistical mechanics and biology. Since bacteria can convert chemical energy into directed motion, they provide prime examples of the constituents of active matter whose macroscopic characteristics can differ strongly from the more traditional passive matter that rests in thermal equilibrium with its environment \cite{TonerTu1995,Czirok2000,Toner2005,Marchetti2013,Bricard2013}.  The generation of this motion necessarily breaks time-reversal symmetry (also known as detailed balance) at the microscopic scale, and it is such inherently nonequilibrium processes that are the focus of modern statistical mechanics.

A major theoretical goal in nonequilibrium statistical mechanics is to identify how the Boltzmann distribution of particle configurations generalises beyond equilibrium conditions.  In equilibrium systems, forces derive from a potential, energy is exchanged reversibly with the environment and the probability of a particle configuration is entirely determined by the potential.  In nonequilibrium systems, where energy is exchanged irreversibly with the environment, there is no one-to-one relationship between a potential that governs interparticle forces and the probability distribution, even in a stationary state.  This means that \emph{effective} forces between particles can emerge as a consequence of the microscopic breaking of detailed balance \cite{Cates2012}. 

A canonical example of an emergent nonequilibrium force is an attraction between self-propelled particles that causes them to cluster macroscopically \cite{Fily2012, GRedner2013}. This attraction can be sufficiently strong that clusters form even if the interaction potential is purely repulsive.  This striking phenomenon arises from particle velocities decreasing as the local particle density around them increases, and is known as motility induced phase separation \cite{Cates2015}.  There are a number of different mechanisms that can generate a density-dependent velocity, thereby breaking detailed balance, with the precise form of the density dependence depending on microscopic considerations.

Most obviously, particles can interact by direct collisions, resulting in jamming where both particles stop moving. This may be considered an extreme case of density dependence \cite{GRedner2013, Soto2014, us}. Other possibilities are density-dependent responses induced by chemotaxis \cite{Berg1972} or other signalling molecules \cite{Miller2001}, and hydrodynamic interactions \cite{Lauga2009}. Theoretical investigations using coarse-grained models of self-propelled particles have succeeded in deriving criteria for motility induced phase separation to occur \cite{Tailleur2008,Wittkowski2014,solon2015active}. 
However---because they explicitly leave out the specific details of the microscopic detailed-balance breaking mechanism---these coarse-grained models lack the power to quantify the relationship between the effective attraction that arises between particles and the underlying microscopic dynamics.

In this work, inspired by the fact that motile bacteria can self-organise into complex macroscopic structures through mechanisms unavailable to passive equilibrium matter \cite{Budrene1995,Wu2000,Sokolov2007,Kearns2010,Fu2012}, we investigate the relationship between microscopic dynamics and emergent behaviour in the context of a simple model of bacterial dynamics.  Specifically, we consider the run-and-tumble motion exhibited by certain bacteria (notably \textit{Escherichia coli} \cite{Berg1972}) whereby self-propulsion generates a series of movements in a fixed direction (runs) interspersed by tumbles that cause a new run direction to be chosen. The most idealised model of this process---that of a persistent random walker---comprises a series of straight-line runs at velocity $v$ with tumble events (occurring as a Poisson process at rate $\tilde\alpha$) that immediately randomise a particle's direction \cite{Schnitzer1993}. In one dimension, there are only two possible directions of movement (`left' and `right'), and the randomisation that occurs on tumbling corresponds to one of the two directions being assigned with equal probability. Consequently, the rate of velocity reversal is $\tilde\alpha/2$ (since there is some probability of maintaining the current direction).

In the case of a single particle and a constant velocity $v$, this model coincides with the persistent random walk which is mathematically equivalent to the dynamics of the voltage and current in power transmission lines as modelled by the telegrapher's equations (see e.g.~\cite{Weiss2002}). The single-particle dynamics is now well understood, including generalisations to a spatially-dependent particle velocity or tumbling rate\cite{Schnitzer1993} and consideration of first-passage properties \cite{Angelani2015}. This single-particle description has also been used as the starting point in a coarse-grained many-body theory for interacting run-and-tumble bacteria \cite{Tailleur2008,Cates2012} that is couched in terms of a mesoscopic density field. This allows one to determine conditions under which phase separation into low- and high-density regions may occur.

Understanding at a more microscopic level has been obtained with reference to lattice-based models of the bacterial dynamics \cite{Thompson2011,Soto2014,Sepulveda2016,us}. In these models, space is discretised and particles hop between neighbouring sites on a lattice instead of moving continuously. The simplest interaction rule to implement is hard-core exclusion, whereby no two particles can occupy the same site simultaneously (although softer rules that allow multiple occupancy are sometimes implemented \cite{Thompson2011,Sepulveda2016}). An advantage of discrete models is that the stochastic dynamics can be formulated exactly (and without ambiguity) using a master equation, which serves as a starting point for analysis. Moreover, such models are typically easier to implement in computer simulations than their continuous-space counterparts. Together, these analytical and numerical methods have shown that the coarse-grained many-body theory described above is recovered in the appropriate limit \cite{Thompson2011} and that dependence of the particle hop rate on the local density that is implied by hard-core exclusion leads to the formation of particle clusters \cite{Soto2014,Sepulveda2016}. The origin of this effect has recently been postulated to lie in an effective attraction between particles whose form was obtained through an exact solution of the master equation for a pair of interacting persistent random walkers \cite{us}. 

Mathematically tractable models are necessarily highly idealised. Nevertheless, certain aspects of true run-and-tumble bacterial dynamics do appear to be well captured. For example, the assumption that the transition from the running state to the tumbling state is a Poisson process corresponds to an exponential distribution of run lengths, which is apparently characteristic of \textit{E.~Coli} \cite{Saragosti2012}. Meanwhile, although it is most natural to think of bacteria being able to access a two- or three-dimensional environment, their behaviour when confined to one-dimensional channels is of experimental interest \cite{Mannik2009}. However, it is not necessarily the case that velocity randomisation is immediate when a tumble event occurs. Some experimental studies (again for \textit{E.~Coli}, \cite{Saragosti2012}) suggest that the tumbling duration also follows an exponential distribution, suggesting that both entry to and exit from the tumbling state can be modelled as Poisson processes (although other distributions have been suggested \cite{Korobkova2004}).

Here, we generalise the exact solution of \cite{us} to the case where the tumbling state is entered at a rate $\tilde\alpha$ and exited at a rate $\tilde\beta$ thus generalising the persistent random walker to a run-and-tumble random walker. While running, particles travel at fixed speed $v$, and whilst tumbling they are stationary. After tumbling, they adopt one of the two possible directions with equal probability. The addition of a finite tumbling time renders the model more faithful to true run-and-tumble bacterial dynamics
and introduces a wider parameter space within which to study the nonequilibrium state. As in \cite{us}, we find an exact solution for the stationary state of the two-particle system, from which we find the exact form of the emergent effective interactions resulting from mutual exclusion and tumbling dynamics.  The solution turns out to involve a multicomponent generating function that satisfies a matrix equation. The inversion of this equation leads eventually to the stationary state probability distribution through a nontrivial procedure that we set out in detail below.
As in \cite{us}, the expression for the stationary state simplifies considerably in a scaling limit in which the running motion becomes deterministic and only the 
running and tumbling times remain  stochastic.
 Our main finding is that while particle collisions generate an effective attraction on a microscopic lengthscale, finite tumbling times lead to a second attractive force over a macroscopic scale.

The remainder of this article is structured as follows. First,  we provide a non-technical summary of our results: In subsection \ref{subsection: model}, we define the lattice-based model of interacting run-and-tumble random walkers that represents the focus of our work; we then summarise the exact solution of this model and our corresponding analysis in subsection \ref{subsection: summary of results}. We set out the derivation of these results in detail in sections 2--4. This begins in section \ref{m-eqns and GF mat eqn} with the master equations for the stochastic system and which we write as  a matrix equation for generating functions. We then show in section \ref{section:inversion} how to solve the matrix equation and invert the generating functions. In section \ref{section: scaling limit}, we find the exact off-lattice steady state distribution in the limit where continuous space and time is recovered.  Finally, we conclude in section \ref{section: conclusion}.

\subsection{Lattice Model Definition}
\label{subsection: model}

To facilitate a more precise discussion, we now formally define our lattice model of two run-and-tumble random walkers. We consider two particles moving under stochastic dynamics  on a periodic one-dimensional lattice of $L$ sites.  Each particle occupies one lattice site and has an internal velocity state $\sigma_i= 0, +1,-1$. A value $\sigma = \pm 1$ (hereafter denoted simply $+$ or $-$) indicates a direction of motion to right or left respectively;  a value $\sigma_{i}=\tum$ indicates that the particle is in a tumbling state and remains stationary on its site. Due to the translational invariance of the system, a microscopic configuration is fully specified by $1\le n < L$, the distance between the two particles in units of the lattice spacing, and the two particle velocities, $\sigma_1$ and $\sigma_2$. A right-moving particle ($\sigma_i=+$) hops one site to the right with rate $\gamma$\; ; likewise, a left-moving particle ($\sigma_i=-$) hops with rate $\gamma$ to the left. However when the target site is occupied by another particle,  hopping is not allowed: this implements the hard-core exclusion interaction.

When a particle enters a tumbling state $\sigma_i=0$, the particle stops hopping. The run lengths and tumble durations are both  Poisson-distributed, with rate parameters $\tilde \alpha$ and $\tilde \beta$ respectively: the particle enters the tumbling state from a running state with rate $\tilde \alpha$ and re-enters  a running state from the  tumbling state  with rate $\tilde \beta$.  In the following we shall consider scaled tumbling rates  
defined as $\alpha = \tilde \alpha/\gamma$ and  $\beta = \tilde \beta/\gamma$, i.e., the bare rates rescaled by the particle hopping rate $\gamma$.

\begin{figure}[h]
  \centering
    \includegraphics[width=0.8\linewidth]{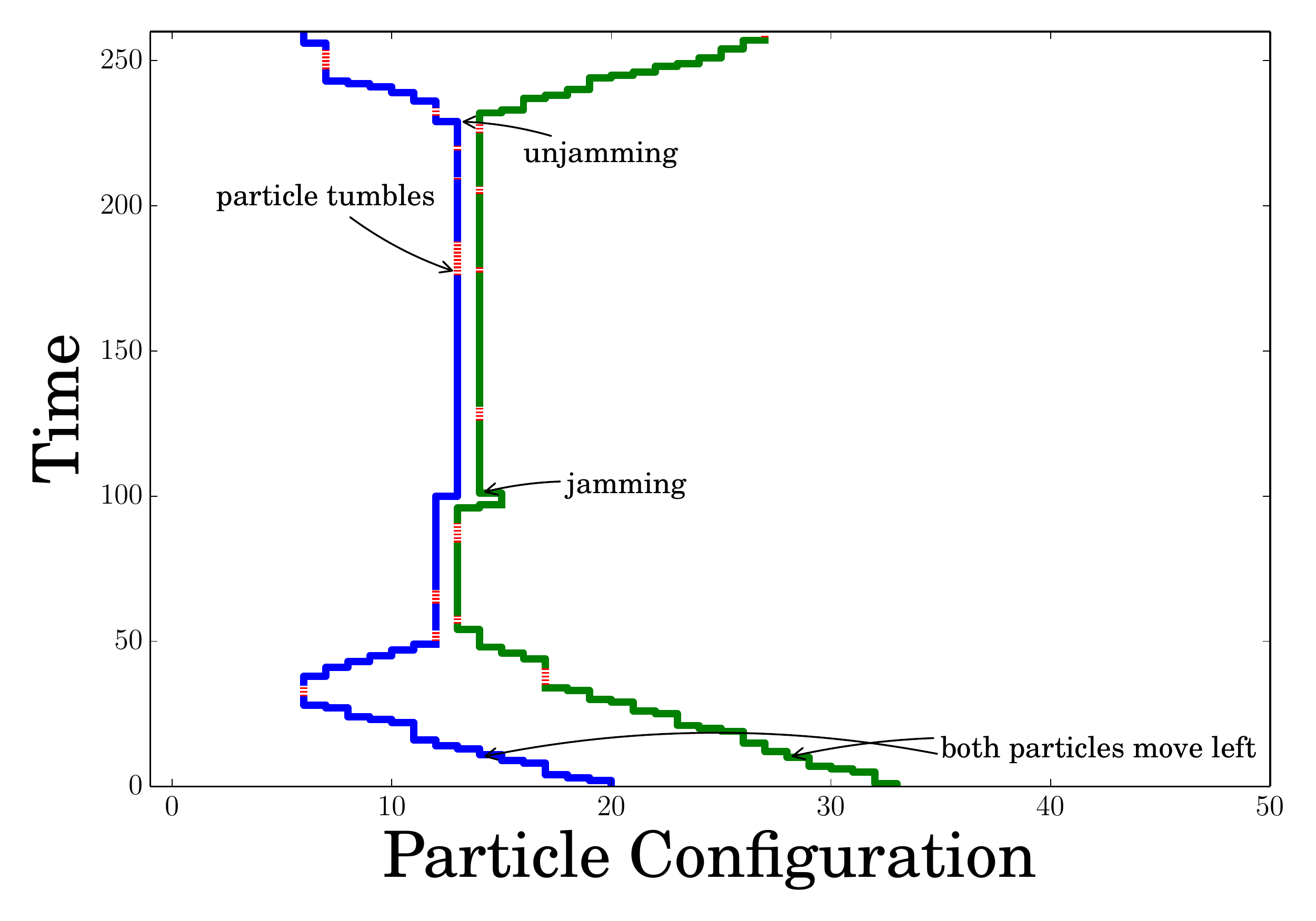}
    \caption{\label{fig:model-ap1-bp9}  Spatiotemporal plot, with time on the $y$ axis, of a simulation of the lattice-based model with $\alpha=0.1$, $\beta=0.9$ and $L=50$. Each line represents a trajectore  of the particle, where particles in the tumbling state are represented by dashed lines.}
\end{figure}

\subsection{Summary of results}
\label{subsection: summary of results}

In this paper, we exactly solve the above lattice model in the steady state. As already mentioned,
each configuration of our model is uniquely described by the distance, $n$, between the two particles and their respective internal velocity states $\sigma_{1},\sigma_{2}$. We find that the exact form for the probability distribution in the steady state, $P_{\sigma_{1}\sigma_{2}}(n) $, is
\begin{eqnarray}
\fl P_{\sigma_{1}\sigma_{2}}(n) &= a_{\sigma_{1}\sigma_{2}}(1) + a_{\sigma_{1}\sigma_{2}}(z_+) z_{+}^{-n+1} + a_{\sigma_{1}\sigma_{2}}(1/z_+) z_{+}^{n-1}  \nonumber \\
\fl &\quad+  a_{\sigma_{1}\sigma_{2}}(z_-) z_{-}^{-n+1} + a_{\sigma_{1}\sigma_{2}}(1/z_-) z_{-}^{n-1} +    w_{\sigma_{1}\sigma_{2}}^{(0)} \delta_{n,1} + w_{\sigma_{1}\sigma_{2}}^{(1)} \delta_{n,L-1} \;,
\label{P intro}
\end{eqnarray}
where the amplitudes $a_{\sigma_{1}\sigma_{2}}(z)$, $w_{\sigma_{1}\sigma_{2}}^{(0)}$ and $w_{\sigma_{1}\sigma_{2}}^{(1)}$, the factors $z_{+}$ and $z_{-}$ are functions of the model parameters $\alpha$, $\beta$ and $L$ and $\delta_{n,m}$ is the Kronecker delta symbol.  In other words, this distribution comprises a constant part and terms that vary exponentially with the particle separation $n$ with further contributions in states where the two particles are next to each other ($n=1$ or $n=L-1$).

We may understand this distribution by considering the dynamics of the jamming that occurs between the interacting run-and-tumble particles. Two particles with equal and opposite velocities collide so that the two particles are on neighbouring sites in either the $(+-, n=1)$ configuration or its symmetrically related counterpart $(-+, n=L-1)$. This results in a microscopically jammed configuration as the particles cannot hop freely until one of them changes orientation. Furthermore, the system cannot change its configuration until one particle starts tumbling. This waiting time is reflected by a delta symbol in the probability of these jammed configurations i.e. $w_{\sigma_{1}\sigma_{2}}^{(0)}$ is nonzero in $P_{+-}(1) $, and $w_{-+}^{(1)}$ is nonzero in $P_{-+}(L-1) $. The second part of the interaction involves unjamming. Eventually the system
enters a jammed configuration of type $(+\tum, n=1)$ or one of the symmetric counterparts,
in which  one of the two particles involved in the collision has begun tumbling.  These configurations therefore also each contain a delta symbol in their probabilities. There is also an enhanced probability of entering the $(\tum \tum, n=1)$ state in which both adjacent particles are tumbling  from these jammed tumbling configurations, which in turn generates delta symbols in $P_{\tum \tum}(1)$, so that $w_{\tum\tum}^{(0)}$ and $w_{\tum\tum}^{(1)}$ are non zero. 
On the other hand $w_{++}^{(0,1)}$ and $w_{--}^{(0,1)}$ are all zero, as there is are no delta-symbol contributions to the probabilities in these velocity sectors. 

Particles unjam by leaving a jammed tumbling configuration; that is when the tumbling particle exits tumbling with an orientation different from the one it had on collision. At this point both particles move in the same direction. Due to the stochasticity of the hopping between lattice sites, some broadening of the separation between the two particles will occur, even though they started off next to each other. This broadening generates a spatially  decaying component in the probability with exponential decay length scale $1/\ln z_{+}(\alpha, \beta)$, that is a function of the tumbling rates $\alpha$ and $\beta$, and is apparent in {\em all} the velocity sectors. Note that in an equilibrium system, an exponentially decaying probability arises from a linear potential with a positive gradient, or---equivalently---a constant attractive force. In this nonequilibrium system, such forces emerge from irreversible collisions between particles. 

Finally, if one of the particles enters a tumbling state when the particles are separated and freely moving,  the result is a configuration where one particle is stationary and tumbling and the other is hopping freely. The freely moving particle either hops towards or away from the tumbling particle. This contribution to the stationary probability distribution is characterised by an exponential decay, but with a new  length scale $ 1/ \ln z_{-}(\alpha,\beta)$. This completes our discussion of the different microscopic mechanisms that lead to (\ref{P intro}).

As noted in the introduction, this lattice-based model is an approximation to the real-world situation of continuous space and time. In order to recover continuum dynamics, we take the lattice spacing as  $\ell/L$  where $\ell$ is the physical system size and let $L \to \infty$. In order to keep the physical velocity
\begin{equation}
v = \gamma \frac{\ell}{L}
\label{vdef}
\end{equation}
invariant we also scale the hopping rate $\gamma$ with system size
\begin{equation}
\gamma =   L/\ell\;.
\label{gscale}
\end{equation}
Then $v =  1$ and
the scaled tumbling rates (ratio of  bare  tumbling rates to hopping rates)
scale as $1/L$:
\begin{eqnarray}
\alpha = \frac{\tilde \alpha}{\gamma}  = \frac{\phi}{L} \hspace{1cm} \beta = \frac{\tilde \beta}{\gamma} = \frac{\theta}{L}
\label{scaling limit def}
\end{eqnarray}
where $\phi = \ell \tilde \alpha$ and $\theta= \ell \tilde \beta$ are dimensionless constants.
Thus in this {\em scaling  limit} the particles undergo a ballistic motion
with velocity $v=1$ interrupted by collision events and stochastic tumbles.

In the scaling  limit, we find the steady-state probabilities of walkers at a separation $y$ have the following form:
\begin{eqnarray}
P_{++}(y) & = a_{++} +  b_{++}[\delta(y) + \delta(\ell-y) ] + c_{++}[e^{- y/\xi} + e^{-(\ell-y)/\xi} ]  \label{++ scaling limit}  \\
P_{+-}(y) & = a_{+-} + c^{(0)}_{+-} e^{- y/\xi} + c^{(1)}_{+-} e^{- (\ell-y)/\xi}  +  w_{+-}\delta(y) \\
P_{+\tum}(y) & = a_{+\tum} + c^{(0)}_{+\tum} e^{- y/\xi } + c^{(1)}_{+\tum}e^{- (\ell-y)/\xi} + w_{+\tum}\delta(y) \\
P_{\tum \tum}(y)  & = a_{\tum \tum} + c_{\tum \tum}[e^{- y/\xi} + e^{-(\ell - y)/\xi} ] +  w_{\tum \tum}[\delta(y) + \delta(\ell - y)  ] \label{tum tum scaling limit}
\end{eqnarray}
where  the length scale $\xi$ is given by
\begin{equation}
\label{xi}
\xi  =  \ell \left( \frac{2}{(\theta +\phi ) (\theta +2 \phi )}\right)^{1/2} \;.
\end{equation}

The amplitudes $a_{\sigma_{1}\sigma_{2}}$, $b_{\sigma_{1}\sigma_{2}}$ and $c_{\sigma_{1}\sigma_{2}}$ derive from the amplitudes $a_{\sigma_{1}\sigma_{2}}(z)$ that appear in (\ref{P intro}). They are functions of the dimensionless parameters $\theta$ and $\phi$, and are specified explicitly in section \ref{section: scaling limit}. Superscripts appear where these amplitudes are different at separation $y=0$ and $y=\ell$. 

A feature of this scaling limit is that in the $++$ (and symmetric $--$) sector the terms containing $z_{+}$ to some power have become delta functions. Therefore they have gone from a finite length scale $1/ \ln z_{+}$ to a delta-function one. The origin of this vanishing lengthscale lies in the fact that the runs are no longer described by stochastic hops, which in the lattice model led to broadening of the particle separation. In the other sectors the $z_{+}$ terms disappear for the same reason. The second length scale $ 1/\ln z_{-}$,  however, remains finite and present in all velocity sectors in the scaling limit
resulting in the decay length $\displaystyle \xi = \frac{\ell}{L} \frac{1}{\ln z_-}$ (\ref{xi}).  This is because the tumble duration---and hence, distance travelled by a moving particle when the other particle is tumbling---remains finite in this limit.

Equations (\ref{++ scaling limit}--\ref{tum tum scaling limit}) are the main results of this work, and demonstrate the rich structure that nonequilibrium steady states may have in comparison to their equilibrium counterparts. The rest of this paper sets out the derivation of these results.

\section{Master equations and generating function matrix equation}
\label{m-eqns and GF mat eqn}

As our model is a Markov process, it can be couched as a system of master equations. We seek the stationary probability distribution of configurations, which are specified in terms of the two particle velocities, $\sigma_{1}\sigma_{2}$, where $\sigma_i = \{ +1, 0, -1\}$, and the particle separation $n$. There are nine velocity sectors in our model:
 ${P}_{++}(n)$,
 ${P}_{+-}(n)$,
 ${P}_{-+}(n)$,
 ${P}_{--}(n)$,
 ${P}_{\tum +}(n)$, ${P}_{+ \tum}(n)$,
 ${P}_{\tum -}(n)$,
 ${P}_{- \tum}(n)$, and
 ${P}_{\tum \tum}(n)$. 
 The symmetry relations between the states due to the periodic boundary conditions and direction-inversion symmetry are as follows:
$
 P_{++}(n) = P_{--}(n), 
 P_{+\tum}(n) = P_{\tum -}(n),
 P_{\tum +}(n) = P_{- \tum}(n),
 P_{+-}(n) = P_{-+}(L-n), 
 P_{+\tum}(n) = P_{\tum +}(L-n), 
 P_{-\tum}(n) = P_{\tum -}(L-n)$.
 Due to these symmetry relations, only the $(++)$, $(+-)$,  $(-+)$, $ (\tum +) $, $ (+ \tum) $, and $ (\tum \tum) $ sectors are independent. The master equations for these velocity sectors are as follows (recalling
$\alpha = \tilde \alpha/\gamma$ and $\beta = \tilde \beta/\gamma$ are scaled rates)
\begin{eqnarray}
\gamma^{-1}\dot{P}_{++}(n)   &= P_{++}(n-1)I_{n>1} + P_{++}(n+1)I_{L-n>1} + \frac{\beta}{2} [P_{\tum +}(n) + P_{+ \tum}(n)]\nonumber \\
		       	   &\quad - P_{++}(n) [ I_{n>1} + I_{L-n>1} + 2\alpha] \label{P++}\\
\gamma^{-1}\dot{P}_{+-}(n)   &= 2P_{+-}(n+1)I_{L-n>1}  + \frac{\beta}{2} [P_{\tum -}(n) + P_{+ \tum}(n)]  \nonumber \\
		           &\quad - P_{+-}(n) [ 2I_{n>1} + 2\alpha ] \\
\gamma^{-1}\dot{P}_{-+}(n)   &= 2P_{-+}(n-1)I_{n>1} + \frac{\beta}{2} [P_{\tum +}(n) + P_{- \tum}(n)]   \nonumber \\
		           &\quad - P_{-+}(n) [ 2I_{L-n>1} + 2\alpha ] \\
\gamma^{-1}\dot{P}_{\tum +}(n) &= P_{\tum +}(n-1)I_{n>1}  + \alpha [ P_{++}(n) + P_{-+}(n) ] + (\beta/2)P_{\tum \tum}(n) \nonumber \\ 
				&\quad - P_{\tum +}(n) \left[  I_{L-n>1} + \alpha + \beta  \right] \\
\gamma^{-1}\dot{P}_{+ \tum}(n) &= P_{+ \tum}(n+1)I_{L-n>1} + \alpha [ P_{++}(n) + P_{+-}(n) ] + (\beta/2)P_{\tum \tum}(n) \nonumber \\
				&\quad - P_{+ \tum}(n) \left[  I_{n>1} + \alpha+ \beta   \right] \\
\gamma^{-1}\dot{P}_{\tum \tum}(n) &= \alpha [ P_{+ \tum}(n) + P_{\tum +}(n) + P_{- \tum}(n) + P_{\tum -}(n) ] - 2\beta P_{\tum \tum}(n)	
\label{P00}
\end{eqnarray}
where  the dot  denotes time derivative. In these equations the indicator $I_{k>1} = 1$ if $k > 1$ and is zero otherwise. The stationary solution satisfies $\dot{P}_{\sigma_1\sigma_2}(n) = 0$ in all sectors.

To find the stationary solution, we introduce the generating functions 
\begin{equation}
G_{\sigma_{1}\sigma_{2}}(x) = \sum_{n=1}^{L-1}x^{n} P_{\sigma_{1}\sigma_{2}}(n)
\label{Gdef}
\end{equation}
and transform the  master equations (\ref{P++}--\ref{P00}) into a system of equations for $G_{\sigma_{1}\sigma_{2}}(x)$. 

For illustrative purposes, let us work through the transformation of the equation for $\dot{P}_{++}(x)$ (\ref{P++}) explicitly as an example.
Summing (\ref{P++}) gives 
 the time evolution of $\dot G_{++}(x)= \sum_{n=1}^{L-1}x^{n} \dot{P}_{++}(n)$, 
\begin{eqnarray}
\fl \gamma^{-1} \dot{G}_{++}(x)&=& \sum_{n=1}^{L-1}x^{n} \bigg( P_{++}(n-1)I_{n>1}  + P_{++}(n+1)I_{L-n>1} + \frac{\beta}{2} [P_{\tum +}(n) + P_{+ \tum}(n)] \nonumber \\ 
\fl			&&  - P_{++}(n) [ I_{n>1} + I_{L-n>1} + 2\alpha]  \bigg)\\
\fl & = &xG_{++}(x)   - x^{L}P_{++}(L-1) 
+\frac{1}{x}G_{++}(x)   - P_{++}(1) 
+\frac{\beta}{2} [ G_{\tum +}(x) + G_{+\tum}(x) ]\nonumber  \\
\fl &&  -G_{++}(x)   + x P_{++}(1 ) 
-G_{++}(x)   + x^{L-1} P_{++}(L-1)  -2 \alpha G_{++}(x) \\
\fl &=  &  [x + x^{-1} - (2+2\alpha) ]G_{++}(x)   - [x^{L}P_{++}(L-1) + P_{++}(1) -xP_{++}(1) ] \nonumber \\
\fl &			   & + x^{L-1}P_{++}(L-1) +\frac{\beta}{2} [ G_{\tum +}(x) + G_{+\tum}(x) ].
 \label{eg GF}
\end{eqnarray}
Finally we use the symmetry
$P_{++}(1) = P_{++}(L-1)$ to obtain
\begin{eqnarray}
 \gamma^{-1} \dot{G}_{++}(x)&=&  \left( x + \frac{1}{x}-2(1+\alpha)\right) G_{++}(x) \nonumber \\
&&+ \frac{\beta}{2} [ G_{\tum +}(x) + G_{+\tum}(x) ]
+(x-1)(1-x^{L-1}) P_{++}(1)\;.
\end{eqnarray}

The remaining generating function equations are as follows:
\begin{eqnarray}
\gamma^{-1}  \dot{G}_{+-}(x) &= [2x^{-1} -(2 + 2\alpha) ]G_{+-}(x) -[2P_{+-}(1) - 2xP_{+-}(1)]  \nonumber \\ \fl &+\frac{\beta}{2} [ G_{\tum -}(x) + G_{+ \tum}(x) ]\\
\gamma^{-1}  \dot{G}_{-+}(x) &= [2x -(2 + 2\alpha) ]G_{-+}(x) -2[x^{L}P_{-+}(L-1) -x^{L-1}P_{-+}(L-1) ]   \nonumber \\
		       &+\frac{\beta}{2} [ G_{\tum +}(x) + G_{- \tum}(x) ] \\
\gamma^{-1}  \dot{G}_{\tum +}(x) &=  [x - (1 + \alpha + \beta) ]G_{\tum +}(x) - [x^{L}P_{\tum _+}(L-1) -x^{L-1}P_{\tum _+}(L-1)]  \nonumber \\
				&+ \alpha [G_{++}(x) + G_{-+}(x)]  + (\beta/2)G_{\tum \tum}(x) \\
\gamma^{-1}  \dot{G}_{+ \tum}(x) &=  [x^{-1} - (1 + \alpha + \beta) ]G_{+ \tum}(x) -(1-x)P_{+ \tum}(1) \nonumber \\
				&+ \alpha [G_{++}(x) + G_{+-}(x)] + (\beta/2)G_{\tum \tum}(x) \\
\gamma^{-1}  \dot{G}_{\tum \tum}(x) &= \alpha[ G_{+ \tum}(x) + G_{\tum +}(x) + G_{- \tum}(x) + G_{\tum -}(x) ] -2\beta G_{\tum \tum}(x).
\end{eqnarray}
We can  close the system
 by making use of the symmetries $G_{\tum +}(x) = G_{- \tum}(x) $ and $G_{+ \tum}(x) = G_{\tum -}(x)$. Similarly, the number of undetermined constants, such as $P_{+-}(1)$ and $P_{-+}(L-1)$, that appear on the right-hand side can be reduced to just three, namely $P_{++}(1)$, $P_{+-}(1)$ and $P_{+\tum}(1)$, by using the symmetries $P_{++}(L-1) = P_{++}(1) $, $P_{-+}(L-1) = P_{+-}(1)$, and $P_{\tum +}(L-1) = P_{+\tum}(1) $. The stationarity condition $\dot{P}_{\sigma_{1}\sigma_{2}}(n) = 0$ translates to $\dot{G}_{\sigma_{1}\sigma_{2}}(x) = 0$ for the generating functions.

After imposing the stationarity condition, we can write this system of equations as a matrix equation in which all the generating functions appear on one side, and all the boundary conditions on the other side. This reads 
\begin{equation}
A(x)\underline{G}(x) = (1-x)\underline{b}(x)
\label{matrix equation}
\end{equation}
  where 
\begin{eqnarray}
A(x) = \left( \begin{array}{c c c c c c}
\mu(x) + \nu(x) & 0 & 0 & \beta/2 & \beta/2 & 0 \\
0 & \nu(x) & 0 & 0 & \beta/2  & 0\\
0 & 0 & \mu(x) & \beta/2  & 0 & 0\\
\alpha & 0 & \alpha & \mu(x) - \beta & 0 & \beta/2\\
\alpha & \alpha & 0 & 0 & \nu(x) - \beta & \beta/2\\
0 & 0 & 0 & \alpha & \alpha & -\beta
\end{array} \right),
\label{matG}
\end{eqnarray}
\begin{eqnarray}
\underline{G}(x) =
\left( \begin{array}{c}
G_{++}(x) \\ G_{+-}(x) \\ G_{-+}(x) \\ G_{\tum +}(x) \\ G_{+ \tum}(x) \\ G_{\tum \tum}(x)
\end{array} \right),\qquad
\underline{b}(x) =
\left( \begin{array}{c}
(1- x^{L-1}) P_{++}(1) \\ P_{+-}(1) \\ -x^{L-1}P_{+-}(1) \\ -x^{L-1}P_{+ \tum}(1)  \\ P_{+ \tum}(1) \\ 0
\end{array} \right), \label{bdef}\\ \nonumber 
\end{eqnarray}
and
\begin{equation}
\mu (x) = x - (1+ \alpha) \;\;\; \textrm{and} \;\;\; \nu(x) = x^{-1} - (1+\alpha) = \mu(x^{-1}).\label{mudef}
\end{equation}

\section{Inversion: a power counting strategy}
\label{section:inversion}

We solve the matrix equation (\ref{matrix equation}) for the generating function vector $\underline{G}(x)$ by inversion:
\begin{eqnarray}
\underline{G}(x) = (1-x) A^{-1}(x)  \underline{b}(x) .
\label{compact matrix equation}
\end{eqnarray}
Our aim is to write the generating functions $G_{\sigma_{1}\sigma_{2}}(x)$ in a  form which allows the probabilities to be read off as coefficients of a power series in $x$. With this in mind, we find that the most convenient form of each generating function is
\begin{equation}
G_{\sigma_{1}\sigma_{2}}(x) = \sum_{\rho} \left[ \frac{xM_{\sigma_{1}\sigma_{2},\rho}}{(1-x/z_{\rho})} \right] + w_{\sigma_{1}\sigma_{2}}^{(0)}x + w_{\sigma_{1}\sigma_{2}}^{(1)}x^{L-1}  + H_{\sigma_{1}\sigma_{2}}(x),
\label{pdGi}
\end{equation}
where $M_{\sigma_{1}\sigma_{2},\rho}$ and $w_{\sigma_{1}\sigma_{2}}$ are functions of the model parameters $\alpha, \beta$ and $L$ (but independent of $x$), $H_{\sigma_{1}\sigma_{2}}(x)$ are polynomials of order greater than $x^{L-1}$, and ${\rho}$ labels the roots $z_{\rho}$ of the determinant of the matrix $A$. 

The stationary probabilities can be read off very quickly as the coefficients of $x^{n}$ by rewriting each fraction $\frac{xM_{\sigma_{1}\sigma_{2},\rho}}{(1-x/z_{\rho})}$ in (\ref{pdGi}) as a geometric series $\sum_{n=1}^{L-1} [\sum_{\rho} M_{\sigma_{1}\sigma_{2},\rho}z_{\rho}^{-n+1}x^{n}] + O(x^{L}) $. We find
\begin{equation}
P_{\sigma_{1}\sigma_{2}}(n) = \sum_{\rho} M_{\sigma_{1}\sigma_{2},\rho}z_{\rho}^{-n+1} + w_{\sigma_{1}\sigma_{2}}^{(0)} \delta_{n,1} + w_{\sigma_{1}\sigma_{2}}^{(1)} \delta_{L-1,1} \;,
\label{general P}
\end{equation}
since terms of order greater than $x^{L-1}$ in (\ref{pdGi}) do not contribute to the probability distribution: the separation $n$ only goes up to $L-1$. (In fact, all terms of degree greater than $x^{L-1}$ will cancel out as the generating functions we have introduced (\ref{Gdef})
do not contain terms at that order.)

In order to obtain the form (\ref{pdGi}) for 
$G_{\sigma_{1}\sigma_{2}}(x)$, we first re-write (\ref{compact matrix equation}) in terms of the adjugate of $A$ (which is defined as the transpose of the cofactor matrix),  the determinant of $A$ and the vector $\underline{b}$ as follows
 \begin{equation}
 \fl G_{\sigma_{1}\sigma_{2}}(x) = (1-x)\sum_{j=1}^{6} A^{-1}_{\sigma_{1}\sigma_{2},j}(x) b_{j}(x) = (1-x)\frac{\sum_{j} \adj A_{\sigma_{1}\sigma_{2},j}(x)b_{j}(x)}{\det A(x)},
 \label{adj Gi}
 \end{equation}
 where the $\sigma_{1}\sigma_{2}$ subscript of $A^{-1}$ indicates the row of $A^{-1}$ that corresponds to that generating function, e.g.~$++$ corresponds to the first row of $A^{-1}$; $j$ is the column number of $A^{-1}$, and $\adj A$ is the adjugate of $A$.

An explicit expression for $\det A(x)$ is
\begin{eqnarray}
\label{detA}
\fl \quad {} \det A(x) =
 &-&\frac{ \beta (2+\alpha) (x-1)^2}{4x^3}  \Big\{ \\ \nonumber
&&   2 (1+ \alpha +\beta) \\\nonumber
 &+& \left[ \alpha  \beta -2 (\alpha +\beta ) (3 \alpha +\beta )-4 (3
   \alpha +2 \beta +2) \right] x \\\nonumber
&+& 2 \left[ (\alpha +\beta ) \left(2 \alpha ^2+\alpha  \beta +6
   \alpha +2 \beta \right)+(\alpha -6) (10-\beta )+66 \right]x^{2} \\ \nonumber
&+&   \left[ \alpha  \beta -2 (\alpha +\beta ) (3 \alpha +\beta )-4 (3
   \alpha +2 \beta +2) \right]x^{3} \\ \nonumber
&+ &   2 (1+ \alpha +\beta)x^{4}     \Big\} \;. 
\end{eqnarray}
 
\subsection{The determinant as a rational function}

To arrive at the form (\ref{pdGi}), we first note that the determinant of $A$, (\ref{detA}), can be written as the following polynomial fraction 
\begin{equation}
 \frac{\det A(x)}{1-x} = \frac{k}{x^{3}}q(x)
\label{det A factorised}
\end{equation}
where
\begin{equation}
q(x) \equiv (x-1)(x-z_{+})(x-1/z_{+})(x-z_{-})(x-1/z_{-}),
\label{qdef}
\end{equation} 
and  
\begin{equation}
k =\frac{\beta}{2} (2+\alpha)(1+\alpha +\beta)\label{kdef}
\end{equation} 
is a constant. 
In expression (\ref{qdef}), $z_{+}$ and $z_{-}$ are independent roots of the determinant, and $1/z_{+}$ and $1/z_{-}$ are their inverses. There is also a root at $z=1$. This furnishes the five roots $z_\rho$ that appear in (\ref{pdGi}). To be explicit: $z_0 = 1$, $z_1 = z_{+}$, $z_2=1/z_{+}$, $z_3=z_{-}$ and $z_4=1/z_{-}$.

That this factorisation of the determinant holds can be seen from (\ref{detA}), where the term in braces is a symmetric quartic polynomial in which the coefficient of the leading term is $2(1+\alpha+\beta)$. The roots come in these reciprocal pairs due to the symmetry of this polynomial.

\subsection{The generating function as a sum of rational functions}

We now manipulate the expression in (\ref{adj Gi}) with a view to writing it in the form of (\ref{pdGi}). Given that we may rewrite $\det A$ as a polynomial fraction (\ref{det A factorised}, \ref{qdef}), we write (\ref{adj Gi}) as
\begin{eqnarray}
 G_{\sigma_{1}\sigma_{2}} &= \frac{-x^{3}(1-x)\sum_{j} \adj A_{\sigma_{1}\sigma_{2},j}(x)b_{j}(x)}{k(x-1)^2(x-z_{+})(x-1/z_{+})(x-z_{-})(x-1/z_{-})} \nonumber \\[1ex]
 &= \frac{x^{3}\sum_{j} \adj A_{\sigma_{1}\sigma_{2},j}(x)b_{j}(x)}{kq(x)} \cdot
\label{Gadic}
\end{eqnarray}
We may separate terms in $G_{\sigma_{1}\sigma_{2}}(x)$ as follows
\begin{eqnarray}
G_{\sigma_{1}\sigma_{2}} = \frac{x^{3}\sum_{j} \adj A_{\sigma_{1}\sigma_{2},j}(x)b_{j}(x)}{kq(x)} =  \frac{x p_{\sigma_{1}\sigma_{2}}(x)}{q(x)} + \frac{\tilde H_{\sigma_{1}\sigma_{2}}(x)}{q(x)},
\label{GF general}
\end{eqnarray}
where each combination $x p_{\sigma_{1}\sigma_{2}}(x)$ is a polynomial of degree less than $x^{L}$ and  $\tilde H_{\sigma_{1}\sigma_{2}}(x)$ is a polynomial with a lowest order term  $x^{L}$. Thus $x p_{\sigma_{1}\sigma_{2}}(x)$  contains all the terms of $x^{3}\sum_{j} \adj A_{\sigma_{1}\sigma_{2},j}(x)b_{j}(x)$ with degree  less than $L$.

We now show that the only terms in $xp_{\sigma_{1}\sigma_{2}}$ of order greater than $x^{5}$ are of order  $x^{6}$ and $x^{L-1}$. 
To do this, we consider  those terms in $x^{3}\adj A_{\sigma_{1}\sigma_{2},j}(x)b_{j}(x)$ that  are  $ O (x^{m})$ where $5 < m < L-1$. Cramer's rule allows us to write 
\begin{equation}
\adj  A_{i,j}(x)b_{j}(x)= \det A_{i}\;,
\end{equation}
where $A_{i}$ is the matrix formed by replacing the $i$-th column of $A$ with $\underline{b}$.
 Then we see that $O(x^{6})$ terms  in $ x^3\det A_{i}$
must come from $\mu^{3}$ terms in  $\det A_i$.  Likewise, the $O(x^{L-1})$ terms in $ x^3\det A_{i}$ must come from multiplying $x^{L-1}$ terms in $\underline{b}$ by $\nu(x)^{3}$ terms in $\det A_{i}$. 
Since $\underline{b}$ only contains terms of order $O(1)$ and $O(x^{L-1})$
one can check that all other terms in $x^3\det A_{i}$
are  $O(x^m)$ where either $m > L-1$ or $m< 6$ .

If $p_{\sigma_{1}\sigma_{2}}(x)$ is of  lower degree than $q(x)$ (i.e. lower order than $x^{5}$), then $p_{\sigma_{1}\sigma_{2}}(x)/q(x)$ will be amenable to partial fraction decomposition, but we have seen that this is not the case in general. We therefore separate each $p_{\sigma_{1}\sigma_{2}}(x)$ into those terms that will allow partial fraction decomposition, and those that will not:
\begin{equation}
\frac{xp_{\sigma_{1}\sigma_{2}}(x)}{q(x)} + 
\frac{\tilde H_{\sigma_{1}\sigma_{2}}(x)}{q(x)} =  x\frac{J_{\sigma_{1}\sigma_{2}}(x)}{q(x)} + \frac{K_{\sigma_{1}\sigma_{2}}(x)}{q(x)} + \frac{\tilde H_{\sigma_{1}\sigma_{2}}'(x)}{q(x)},
\label{powersep}
\end{equation} 
where $J_{\sigma_{1}\sigma_{2}}(x)$ takes terms from $p_{\sigma_{1}\sigma_{2}}(x)$ amenable to partial fraction decomposition (ie. those of order less than $q(x)$) and is therefore a polynomial of order $x^{4}$ or less, and $K_{\sigma_{1}\sigma_{2}}(x)$ takes the higher order terms from $p_{\sigma_{1}\sigma_{2}}(x)$. However, at the same time, we want an expression $K_{\sigma_{1}\sigma_{2}}(x)/q(x)$ that can be cast as a polynomial rather than a rational function so that we can read off its contribution to the probability.  To this end, we define 
\begin{equation}
K_{\sigma_{1}\sigma_{2}}(x)/q(x) \equiv w_{\sigma_{1}\sigma_{2}}^{(0)}x +  w_{\sigma_{1}\sigma_{2}}^{(1)}x^{L-1},
\end{equation}
where $w_{\sigma_{1}\sigma_{2}}^{(0)}$ is  equal to the ratio of the coefficient of the $x^{6}$ term in $xp_{\sigma_{1}\sigma_{2}}(x)$ with the  coefficient of the $x^{0}$ term in $q(x)$   and $w_{\sigma_{1}\sigma_{2}}^{(1)}$ is equal to  the ratio of the coefficient of the $x^{L-1}$ term in $xp_{\sigma_{1}\sigma_{2}}(x)$  with the  coefficient of the $x^{5}$ term in $q(x)$.  In order to factorise $K_{\sigma_{1}\sigma_{2}}(x)$  by $q(x)$, we add to $K_{\sigma_{1}\sigma_{2}}(x)$ any terms required, in addition to the $x^{6}$ and $x^{L-1}$ terms in $xp_{\sigma_{1}\sigma_{2}}(x)$ already present. If these added terms are of degree less than $5$, then we subtract them from $p_{\sigma_{1}\sigma_{2}}(x)$. On the other hand, if the added terms are of degree greater than $L-1$ (recall, we have already shown that there are no further terms between $x^{5}$ and $x^{L}$), we subtract them from $\tilde H_{\sigma_{1}\sigma_{2}}(x)$, which in turn becomes $\tilde H_{\sigma_{1}\sigma_{2}}'(x)$.

\subsection{Partial fraction decomposition using the `cover up' method}
We now return to our expressions $J_{\sigma_{1}\sigma_{2}}(x)$ in (\ref{powersep}), which we know are amenable to partial fraction decomposition. A remarkable simplification occurs 
when we use  Heaviside's `cover-up' method 
for the partial-fraction expansion of a rational function \cite{norman1990},
on the fraction $\frac{J_{\sigma_{1}\sigma_{2}}(x)}{q(x)}$.  The method may be used whenever the denominator of a rational fraction can be factorised into distinct linear factors. We have already shown that $q(x)$ can be written in this form, and that each $J_{\sigma_{1}\sigma_{2}}(x)$ is a polynomial, and therefore the method can be applied to our fraction, which yields
\begin{eqnarray}
\fl \frac{J_{\sigma_{1}\sigma_{2}}(x)}{q(x)} &=&  \frac{J_{\sigma_{1}\sigma_{2}}(x)}{(x-z_{1})(x-z_{2})...(x-z_{n})} \nonumber \\
&=& \frac{J_{\sigma_{1}\sigma_{2}}(z_{1})}{(z_{1}-z_{2})...(z_{1}-z_{n})} \cdot \frac{1}{x-z_{1}} + ... +  \frac{J_{\sigma_{1}\sigma_{2}}(z_{n})}{(z_{n}-z_{1})...(z_{n}-z_{n-1})} \cdot \frac{1}{x-z_{n}}
\end{eqnarray}
where $z_{1},...,z_{n}$ are the roots of $q(x)$. The denominators of each fraction in the resulting decomposition are just the linear factors, as is familiar from normal partial fraction decomposition. The corresponding numerators, $J_{\sigma_{1}\sigma_{2}}(z_{i})$, are found by covering up the factor $x-z_{i}$ in  $\frac{J_{\sigma_{1}\sigma_{2}}(x)}{q(x)}$, and setting $x=z_{i}$ in the rest of the expression.
The terms involving $J_{\sigma_{1}\sigma_{2}}$ are now in the form of $\frac{xM_{\sigma_{1}\sigma_{2}}(z_{\rho})}{(1-z_{\rho}x)}$ of (\ref{pdGi}) and so straightforwardly invertible. We can ignore the expressions within $\tilde H_{\sigma_{1}\sigma_{2}}'(x)$ entirely as they do not contribute. We may therefore write the generating function in general as
\begin{eqnarray}
\fl G_{\sigma_{1}\sigma_{2}} = \sum_{\rho} \left[ \frac{xJ_{\sigma_{1}\sigma_{2}}(z_{\rho})}{[q(x)/(x-z_{\rho})]_{|x=z_{\rho}}}\frac{1}{(x-z_{\rho})} \right] + w_{\sigma_{1}\sigma_{2}}^{(0)}x + w_{\sigma_{1}\sigma_{2}}^{(1)}x^{L-1} + x^{L}\tilde H'_{\sigma_{1}\sigma_{2}}.
\label{general GF form}
\end{eqnarray}
We then write an expression for the steady-state probabilities of the form in (\ref{general P})
\begin{equation}
P_{\sigma_{1}\sigma_{2}}(n) = \sum_{\rho}a_{\sigma_{1}\sigma_{2}}(z_{\rho}) z_{\rho}^{-n+1} + w_{\sigma_{1}\sigma_{2}}^{(0)} \delta_{n,1} + w_{\sigma_{1}\sigma_{2}}^{(1)} \delta_{L-1,1}.
\label{specific P}
\end{equation}
where 
\begin{equation}
a_{\sigma_{1}\sigma_{2}}(z_{\rho}) = \frac{-J_{\sigma_{1}\sigma_{2}}(z_{\rho})}{[q(x)/(x-z_{\rho})]_{|x=z_{j=\rho}}} \cdot
\label{define a}
\end{equation}
Thus we have derived the form of the steady-state probability of our system.

\subsection{Weights in different velocity sectors}

It remains to determine which weights $w_{\sigma_{1}\sigma_{2}}^{(i)}$ are non-zero in their corresponding velocity sectors $\sigma_{1}\sigma_{2}$. We proceed column-by-column in $A$, replacing each with $\underline{b}$. Thanks to the symmetries $G_{+-}(x) = G_{-+}(L-x)$ and $G_{+\tum}(x) = G_{\tum+}(L-x)$, we are only required to solve for four generating functions $G_{++}(x), G_{+-}(x), G_{+\tum}(x)$ and $G_{\tum \tum}(x)$. For $G_{++}$, as a $\mu$ is eliminated (replaced by $b_{1}$), it is not possible to get a $O(x^{6})$ term, nor a $O(x^{L-1})$ term as a $\nu$ is also eliminated. For $G_{+-}(x)$, we have sufficient factors of $\mu$ to get an $O(x^{6})$ term but no diagonal $x^{L-1}$ terms for  $O(x^{L-1})$ terms. For $G_{+ \tum}(x)$, we get an $O(x^{L-1})$ term only for the similar reasons. For $G_{\tum \tum}$ both $\mu^{3}$ and $\nu^{3}$ terms are possible, and so $G_{\tum \tum}(x)$ can possess both  $O(x^{6})$ and $O(x^{L-1})$ terms.

\subsection{Determination of the constants}
\label{subsection: lattice consts}

We complete our derivation by briefly describing how to determine  the constants $P_{++}(1), P_{+-}(1)$ and $P_{+\tum}(1)$. We find two of these as yet undetermined constants by imposing the condition that the generating functions must not diverge at any $x$. As $G_{i}$ has poles at each of the roots---the denominator is a product of the linear roots---this condition implies that the numerator,  $\adj A_{i,j}(x)\tilde{b}_{j}(x)$, has to cancel the determinant poles and thus must equal zero at all of the roots. We find that at $x=1$, the numerator is automatically zero. It remains to impose pole cancellation for the roots $z=z_{+},1/z_{+},z_{-}, 1/z_{-}$. Although there are $24$ simultaneous equations following from this condition, we find that only two are linearly independent. Therefore from this condition we can find any two of $P_{++}(1), P_{+-}(1)$ and $P_{+\tum}(1)$. We find the remaining constant by imposing normalisation: $\sum_{\sigma_{1}\sigma_{2}}\sum_{n=1}^{L-1}P_{\sigma_{1}\sigma_{2}}(n)=1$.

\subsection{Plots of the probability distribution}

We have derived the general form of the steady-state probability distribution, (\ref{specific P}) and (\ref{define a}). However, there remain a number of expressions that we have not presented explicitly in terms of the model parameters $\alpha$ and $\beta$, namely the roots $z_{\rho} = z_{\rho}[\alpha,\beta]$, the weights $w_{\sigma_{1}\sigma_{2}}=w_{\sigma_{1}\sigma_{2}}[\alpha,\beta]$, the amplitudes $a_{\sigma_{1}\sigma_{2}}(z_{\rho})=a_{\sigma_{1}\sigma_{2}}(z_{\rho})[\alpha,\beta]$, and the constants $P_{++}(1)=P_{++}(1)[\alpha,\beta]$, $P_{+-}(1)=P_{+-}(1)[\alpha,\beta]$ and $P_{+\tum}(1)=P_{+\tum}(1)[\alpha,\beta]$. It is possible to find these expressions explicitly, but due to their unwieldy form we consign the details to a Mathematica notebook in the  Supplementary Material. The notebook performs an exact analytic calculation of the probability distribution up to the normalisation of the distributions $P_{\sigma_{1}\sigma_{2}}(n)$. Normalisation for a specific set of model parameters is achieved numerically, calculated to arbitrary precision limited only by machine capability.

\begin{figure}[h!]
 \centering
         \includegraphics[width=0.95\textwidth]{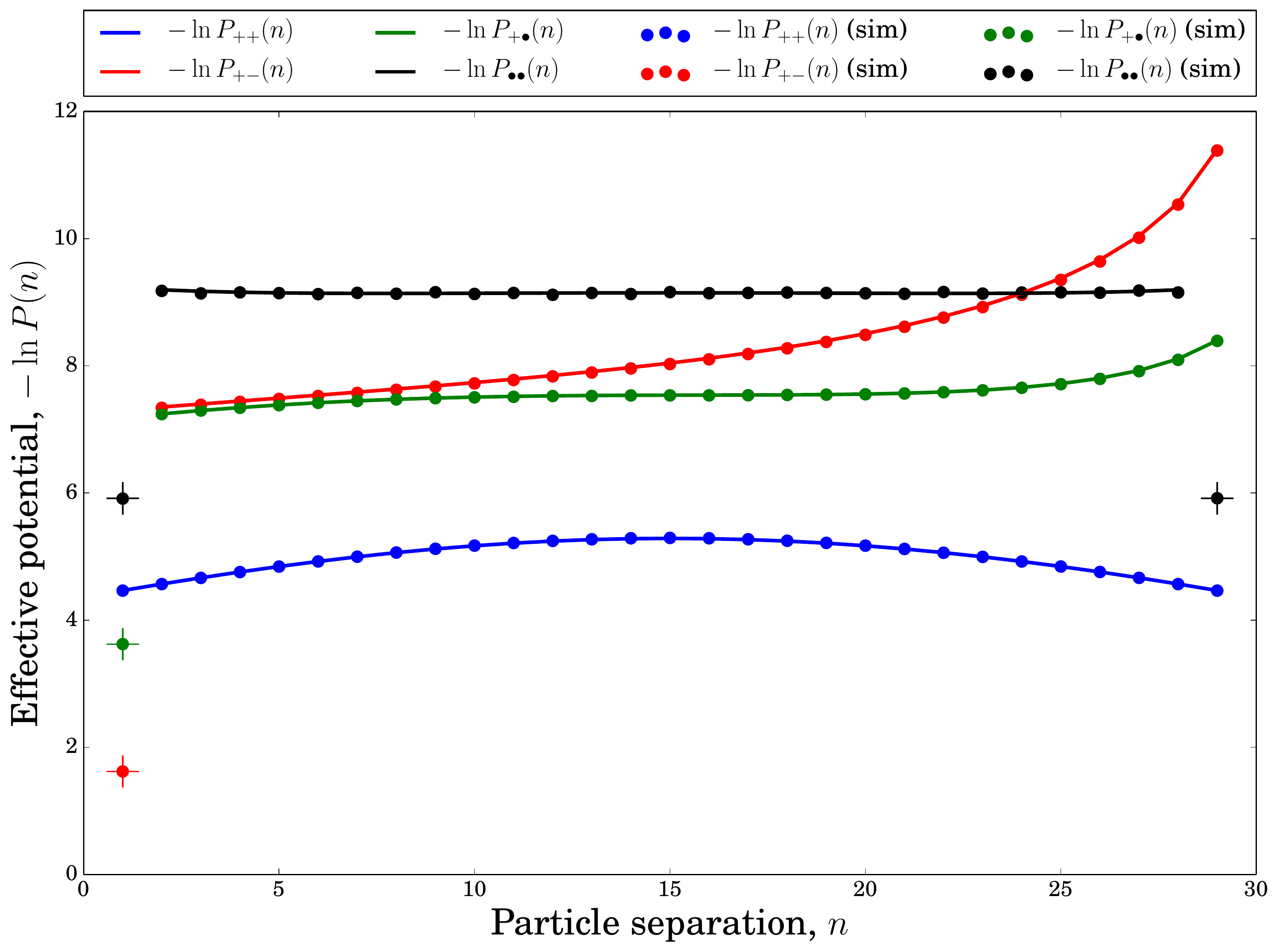}
     \caption{Comparison of analytic calculation of probability and simulation results for $L=30$, $\alpha = 0.01$, $\beta=0.1$. The crosses mark the analytic results that contain delta symbols.}
    \label{lattice model prob}
\end{figure}

A comparison of a simulation with our analytic solution for particular values of the model parameters is shown in Figure~\ref{lattice model prob}, showing complete agreement. As in \cite{us}, we present the results in the form of effective potentials, $V(x) = -\ln P(x)$. Recall that for equilibrium systems, we would obtain a Boltzmann distribution $P \propto {\rm e}^{-V(x)}$. The effective potential for a nonequilibrium tells us what kind of potential an equilibrium system, without internal propulsion, would have to have in order to see the same macroscopic physics.  For simplicity, we plot only the four independent sectors, in which the particles are approaching ($+-$ and $+\tum$ sectors) or maintain a constant (average) separation ($++$ and $\tum\tum$ sectors).  We see there is an attraction towards low separations, $n\ll L$, in the sectors where the particles are approaching, and that the characteristic lengthscales differ between these two approaching sectors.  

\section{Scaling limit of the probability distribution}
\label{section: scaling limit}

Tractable closed-form expressions for the probability distributions can be found in the scaling limit defined by Eqs.~(\ref{vdef})--(\ref{scaling limit def}). To recap briefly, the limit of continuous space is reached by taking the  lattice spacing to zero as $1/L$ while leaving the physical system size $\ell$ fixed. At the same time the hopping rate diverges with $L$ via (\ref{gscale}) in order to leave the physical velocity fixed as $v=1$. 
The resulting limit for the ratio of bare tumbling rates to hopping rate is
\begin{eqnarray}
\alpha = \frac{\tilde \alpha}{\gamma}  = \frac{\phi}{L} \;,\quad \beta = \frac{\tilde \beta}{\gamma} = \frac{\theta}{L} \quad\mbox{and}\quad L \rightarrow \infty
\label{Why use a short label when you can write a long one?}
\end{eqnarray}
with $\phi$ and $\theta$  both constant. In this limit, the running of the bacteria becomes ballistic while tumbling remains stochastic. We have already summarised the form of the stationary  probability (\ref{++ scaling limit})--(\ref{tum tum scaling limit}) in this limit in section \ref{subsection: summary of results}. In this section, we show how to derive exact expressions for  the various amplitudes $a_{\sigma_1 \sigma_2}$, $b_{\sigma_1 \sigma_2}$, $c_{\sigma_1 \sigma_2}$, $w_{\sigma_1 \sigma_2}$ involved. These exact expressions are  written out explicitly in Sec.~\ref{section: scal limit results} for reference.

The derivation of (\ref{++ scaling limit})--(\ref{tum tum scaling limit}) proceeds in three parts. We first find the roots $z_{\rho}$ and the  constants $P_{++}(1)$ and $P_{+-}(1)$ in the scaling limit in terms of $P_{+\tum}(1)$, $\phi$ and $\theta$ by cancelling poles in the generating functions (cf.\ subsection \ref{subsection: lattice consts}). Next, using these expressions, we write the amplitudes in terms of $P_{+\tum}(1)$, $\phi$ and $\theta$ only. Finally, we impose normalisation, which gives $P_{+\tum}(1)$ in terms of $\phi$ and $\theta$ only. At the end of this process we arrive at the normalised probability distribution, $P_{\sigma_{1}\sigma_{2}}(y)$ in terms of the parameters $\phi$, $\theta$ and $\ell$ only.

\subsection{Constants from pole cancellation}

Since each generating function $G_{\sigma_{1}\sigma_{2}}(x)$ is a finite sum (see~(\ref{Gdef})), it cannot diverge at any $x$. Consequently each pole in (\ref{Gadic}), corresponding to a root $z_{\rho}$ of $\det A$, must be cancelled by a zero in the numerator. Each such condition leads to a linear equation in $P_{++}(1)$, $P_{+-}(1)$ and $P_{+\tum}(1)$. As previously noted, it turns out that there are only two linearly independent equations in these quantities. This means that a further condition (namely, normalisation) is required to determine them all.

One of the roots is $z_{\rho}=1$. One can show that the numerator of (\ref{Gadic}) is always zero at $x=1$, which does not provide any information about $P_{++}(1), P_{+-}(1)$ and $P_{+\tum}(1)$. However, at the other roots $x=z_{+},1/z_{+},z_{-}, 1/z_{-}$, the numerator is not automatically zero. Consequently, we impose the condition
\begin{equation}
\sum_{j} \adj A_{\sigma_{1}\sigma_{2},j}(z_{\rho})\,b_{j}(z_{\rho}) = 0,
\label{pole cancellation condition}
\end{equation}
at each of the roots $z_{\rho} = z_{+},z_{-},1/z_{+}$ and $1/z_{-}$. We find the two desired linearly independent conditions by taking $z=z_{+}$ and $z=z_{-}$ in (\ref{pole cancellation condition}) with $\sigma_{1}\sigma_{2}=++$. 
Using the expression (\ref{bdef}) for 
$b_{j}(x) = 0$,
each of these conditions takes the form
\begin{eqnarray}
 \fl \tilde{A}_{++,1}(z_{\rho})P_{++}(1) +  \tilde{A}_{++,2}(z_{\rho})P_{+-}(1) +  \tilde{A}_{++,5}(z_{\rho})P_{+ \tum}(1) \nonumber
\\ =  z_{\rho}^{L-1} \left[ \tilde{A}_{++,1}(z_{\rho})P_{++}(1) + \tilde{A}_{++,3}(z_{\rho})P_{+-}(1) +  \tilde{A}_{++,4}(z_{\rho})P_{+ \tum}(1) \right ]
  \label{root cancelling G++ at z}
\end{eqnarray}
where, for convenience, we introduce the notation $\tilde{A}_{\sigma_{1}\sigma_{2},j}(x) \equiv \adj A_{\sigma_{1}\sigma_{2},j}$. 

To apply these two conditions, we need to know the location of the roots $z_{+}$ and $z_{-}$ of $\det A$, as defined by (\ref{det A factorised}). By expanding $z_{\pm}$ about $1$ in powers of $1/\sqrt{L}$ in the explicit expression (\ref{detA}) for the determinant, one finds that 
\begin{eqnarray}
\label{z+sl}
z_{+} &\sim 1 + \frac{\sqrt{2\phi}}{\sqrt{L}} +\frac{\phi}{L} +O\left(L^{-3/2}\right) \\
z_{-} &\sim 1 + \sqrt{\frac{(\theta +\phi ) (\theta +2 \phi )}{2}} \frac{1}{L} +O\left(L^{-3/2}\right) \;.
\end{eqnarray}
When we substitute these roots into (\ref{root cancelling G++ at z}), we find that $z_{+}^{L-1}\to\infty$, and so the terms in square brackets on the right-hand side of  (\ref{root cancelling G++ at z}) need to cancel at this root. At $z_{-}$, the factor $z_{-}^{L-1}$ approaches ${\rm e}^\lambda$ where
$\lambda = \lim_{L \to\infty} L [ z_{-} - 1 ]$ is  given by
\begin{equation}
\lambda = 
\left(\frac{(\theta +\phi ) (\theta +2 \phi )}{2}\right)^{1/2} \;.
\end{equation}

For future reference, it is also helpful to note the locations of the reciprocal roots 
\begin{eqnarray}
1/z_{+} &\sim 1 - \frac{\sqrt{2\phi}}{\sqrt{L}} +\frac{\phi}{L} +O\left(L^{-3/2}\right) \\
1/z_{-} &\sim 1 - \sqrt{\frac{(\theta +\phi ) (\theta +2 \phi )}{2}} \frac{1}{L} +O\left(L^{-3/2}\right) \;.
\end{eqnarray}

The next step is to determine the leading large-$L$ forms of the adjugate elements $\tilde{A}_{\sigma_{1}\sigma_{2},j}(x)$ appearing in (\ref{root cancelling G++ at z}) at each of the roots. All subleading terms will vanish in the scaling limit.  To identify these leading terms, we require explicit expressions for $\tilde{A}_{\sigma_{1}\sigma_{2},j}(x)$ in the constants $\alpha$, $\beta$ and the functions $\mu(x) = x-(1+\alpha)$ and $\nu(x) = x^{-1}-(1+\alpha) = \mu(x^{-1})$. These can be obtained most straightforwardly using a computational algebra package such as Mathematica. For example, one finds
\begin{eqnarray}
 \tilde{A}_{++,1}(x) &=  -\frac{1}{4} \beta  \bigg(\beta ^2 (\alpha +2 \mu ) (\alpha +2 \nu )-\beta  (\alpha +2 \mu ) (\mu +\nu ) (\alpha +2 \nu ) \nonumber \\ &\quad{}+ 2 \mu  \nu  (\alpha  (\mu
   +\nu )+2 \mu  \nu )\bigg)
\label{A++,1(zpl)}
\end{eqnarray}
where it is important to keep in mind that $\mu$ and $\nu$,
defined in (\ref{mudef}),
 are functions of $x$. Substituting the $L$-dependent expressions for $\alpha$ and $\beta$, (\ref{Why use a short label when you can write a long one?}), along with the large $L$ form of $z_+$ (\ref{z+sl}) into the above expression yields the large-$L$ result
\begin{equation}
\tilde{A}_{++,1}(z_{+}) \sim -\frac{4\theta \phi^{2}}{L^{3}} + O\left( L^{-7/2} \right) \;.
\label{example A}
\end{equation}
Using the same method, one can find the leading terms of each of the adjugate elements in  (\ref{root cancelling G++ at z}) at each of the roots $z_{\rho}$. The results are summarised in Table~\ref{Adjugate elements in the scaling limit for constants}.

\begin{table}
\centering
\caption{Adjugate elements in the scaling limit required to evaluate $P_{++}(1)$ and $P_{+-}(1)$ and $J_{++}$.}
\label{Adjugate elements in the scaling limit for constants}
\begin{tabular}{r|c|c|c}
          & $x=1$ & $x=z_{+}$ & $x=1/z_{-}$  \\ \hline &&&\\[-2ex]
$\tilde{A}_{++,1}$ & $\displaystyle -\frac{\theta ^2 \phi ^2 \zeta}{4 L^5}$     &  $\displaystyle -\frac{4 \theta  \phi ^2}{L^3}$         & $\displaystyle \frac{\theta ^3 \zeta^2}{4 L^5}$                 \\[2ex]
$\tilde{A}_{++,2}$ & $\displaystyle-\frac{\theta ^2 \phi ^2 \zeta}{4 L^5}$     & $\displaystyle-\frac{\theta ^2 \phi ^2}{L^4}$          & $\displaystyle-\frac{\theta ^2 \phi  \zeta \left(2\lambda+\zeta
   \right)}{4 L^5}$              \\[2ex]
$\tilde{A}_{++,3}$ & $\displaystyle-\frac{\theta ^2 \phi ^2 \zeta}{4 L^5} $    & $\displaystyle-\frac{\theta ^2 \phi ^2}{L^4} $         & $-\displaystyle\frac{\theta ^2 \phi  \zeta \left(-2\lambda+\zeta \right)}{4 L^5} $                 \\[2ex]
$\tilde{A}_{++,4}$ & $\displaystyle-\frac{\theta ^2 \phi ^2 \zeta}{4 L^5}  $   & $\displaystyle \frac{\sqrt{2} \theta ^2 \phi ^{3/2}}{L^{7/2}} $         & $\displaystyle-\frac{\theta ^3 \zeta \left(2\lambda-2\zeta \right)}{8 L^5}$                    \\[2ex]
$\tilde{A}_{++,5}$ & $\displaystyle-\frac{\theta ^2 \phi ^2 \zeta}{4 L^5}   $  & $\displaystyle-\frac{\sqrt{2} \theta ^2 \phi ^{3/2}}{L^{7/2}}$         & $\displaystyle\frac{\theta ^3 \zeta \left(2\lambda+2 \zeta \right)}{8 L^5}$                    
\end{tabular}
\end{table} 

Now, solving the two equations arising from substituting $x=z_+$ and $x=1/z_-$ into (\ref{root cancelling G++ at z}) we find for large $L$
\begin{eqnarray}
P_{++}(1) &\sim \frac{\theta}{2 \sqrt{2L\phi }} P_{+\tum}(1) \label{P++(1)}  \\
P_{+-}(1) &\sim  \frac{\theta  \left(e^{\lambda } (\zeta +\lambda )-\zeta +\lambda
   \right)}{\phi  \left(\zeta  \left(e^{\lambda }-1\right)+2 \left(e^{\lambda }+1\right)
   \lambda \right)} P_{+\tum}(1) \label{P+-(1)}
\end{eqnarray}
where
\begin{eqnarray}
\zeta &\equiv \theta + 2\phi  \label{define zeta} \\
\eta   &\equiv \theta + \phi \;. \label{define eta}
\end{eqnarray}
The remaining constant $P_{+\tum}(1)$ will be found by normalisation (see Sec.~\ref{subsection: normalisation} below).

\subsection{Decay lengths and amplitudes}

In the scaling limit, we wish to move from a  
discrete separation of $n$ lattice sites to a continuous separation $y$ that lies between $0$ and $\ell$. This we achieve with the transformation
\begin{equation}
n = \frac{L y}{\ell} \;,
\end{equation}
under which $P_{\sigma_1\sigma_2}(y) = \frac{L}{\ell} P_{\sigma_1\sigma_2}(n)$. The discrete distribution (\ref{general GF form}) contains a set of terms of the form
\begin{equation}
\label{dterm}
a_{\sigma_{1}\sigma_{2}}(z_{\rho}) z_{\rho}^{-n+1} \quad\mbox{where}\quad
a_{\sigma_{1}\sigma_{2}}(z_{\rho}) = \frac{-J_{\sigma_{1}\sigma_{2}}(z_{\rho})}{[q(x)/(x-z_{\rho})]_{|x=z_{j=\rho}}} 
\end{equation}
and $z_{\rho}$ is one of the five roots, $z_{\rho} \in \{ 1, z_+, 1/z_+, z_-, 1/z_-\}$. We now establish their behaviour in the scaling limit.

The easiest case to deal with $z_{\rho}=1$, where the amplitude $a_{\sigma_1\sigma_2}$ that appears in the result for the scaling limit, (\ref{++ scaling limit})--(\ref{tum tum scaling limit}), is equal to $\lim_{L\to\infty}\frac{L}{\ell} a_{\sigma_1\sigma_2}(1)$. 

At $z_{\rho}=z_+$ we have the combination
\begin{equation}
\lim_{L \to \infty} \frac{L}{\ell} a_{\sigma_{1}\sigma_{2}}(z_+) \left(1 + \frac{\sqrt{2\phi}}{\sqrt{L}}\right)^{-n+1} \;,
\end{equation}
which, in terms of the continuous coordinate $y$, becomes
\begin{equation}
\lim_{L \to \infty} \frac{L}{\ell} a_{\sigma_{1}\sigma_{2}}(z_+) \exp( - \sqrt{2\phi L} y ) = \left[ \lim_{L \to \infty} a_{\sigma_{1}\sigma_{2}}(z_+) \sqrt{ \frac{L}{2\phi} } \right] \delta(y)  \;.
\end{equation}
Note that we think of the delta function as being slightly displaced from the boundary at $y=0$, so that the integral $\int_0^{\ell} {\rm d}y\,\delta(y) = 1$. The scaling of $a_{\sigma_{1}\sigma_{2}}(z_+)$ in the large $L$ limit determines whether the delta function actually appears in the $\sigma_1\sigma_2$ sector. In particular, if $a_{\sigma_{1}\sigma_{2}}(z_+)$ decays faster than $1/\sqrt{L}$, we will not get a delta function contribution. The quantity in the square bracket can be identified as $b_{\sigma_1\sigma_2}^{(0)}$ that appears in the probability distribution in the scaling limit. Note that in Equations~(\ref{++ scaling limit})--(\ref{tum tum scaling limit}) we dropped superscripts on the amplitudes in the scaling limit where this was unambiguous.

At $z_\rho = 1/z_+$, we find 
\begin{equation}
 \left[ \lim_{L \to \infty} a_{\sigma_{1}\sigma_{2}}(1/z_+) \sqrt{ \frac{L}{2\phi} } \right] \delta(\ell-y) \;.
\end{equation}
Here, the term in square brackets defines the amplitude $b_{\sigma_1\sigma_2}^{(1)}$.

Turning now to the root $z_{\rho}=z_{-}$, we have
\begin{equation}
\lim_{L \to \infty} \frac{L}{\ell} a_{\sigma_{1}\sigma_{2}}(z_-) \left(1 + \frac{\lambda}{L}\right)^{-n+1} = \left[ \lim_{L\to\infty} \frac{L}{\ell} a_{\sigma_{1}\sigma_{2}}(z_-) \right] \exp\left( - \frac{y}{\xi} \right) 
\end{equation}
in which we have introduced the lengthscale 
\begin{equation}
\xi = \frac{\ell}{\lambda} =  \frac{\sqrt{2} \ell}{\sqrt{(\phi+\theta)(2\phi+\theta)}} \;.
\end{equation}
The square-bracketed term defines the amplitude $c_{\sigma_1\sigma_2}^{(0)}$.

Finally, at $z_{\rho}=1/z_{-}$, we find
\begin{equation}
\lim_{L \to \infty} a_{\sigma_{1}\sigma_{2}}(1/z_-) \left(1 - \frac{\lambda}{L}\right)^{-n+1} = \left[ \lim_{L\to\infty} a_{\sigma_{1}\sigma_{2}}(1/z_-) e^{\lambda} \right] \exp\left( - \frac{\ell-y}{\xi} \right)  \;,
\end{equation}
which furnishes an expression for the amplitude $c_{\sigma_1\sigma_2}^{(1)}$. 

It now remains to evaluate the amplitudes.  Recall that $J_{\sigma_1\sigma_2}(x)$, defined by (\ref{powersep}) is by construction a polynomial of  degree $\leq 4$. Specifically,
\begin{equation}
J_{\sigma_1\sigma_2}(x) = \hat{T}_4 \sum_j x^2 \adj A_{\sigma_1\sigma_2,j}(x) b_j(x)
\end{equation}
where the operator $\hat{T}_4$ discards terms of order $x^5$ and higher in a power series in $x$. This we may write as
\begin{equation}
J_{\sigma_{1}\sigma_{2}}(x) = \tilde{A}_{\sigma_{1}\sigma_{2},1}' P_{++}(1) + \tilde{A}_{\sigma_{1}\sigma_{2},2}'P_{+-}(1) + \tilde{A}_{\sigma_{1}\sigma_{2},5}'P_{+\tum}(1) \;,
\end{equation}
where $\tilde{A}_{\sigma_{1}\sigma_{2},j}'(x) = \hat{T}_4 x^2 \adj A_{\sigma_1\sigma_2,j}(x)$.

In the $++$ sector, the adjugate elements exhibit the symmetries
\begin{eqnarray}
\tilde{A}_{++,1}(x) &= \tilde{A}_{++,1}(1/x) \\
\tilde{A}_{++,2}(x) &= \tilde{A}_{++,3}(1/x) \\
\tilde{A}_{++,5}(x) &= \tilde{A}_{++,4}(1/x) 
\end{eqnarray}
as can be verified by inspection of the explicit expressions presented in the Supplementary Material.  Using these symmetries in (\ref{root cancelling G++ at z}), one can show that
\begin{equation}
J_{++}(1/z_{\rho}) = z_{\rho}^{1-L}J_{++}(z_\rho)
\label{++ symm}
\end{equation}
at each of the roots $z_\rho$. The same symmetry also applies in the $\tum\tum$ sector, namely $J_{\tum\tum}(1/z_{\rho}) = z_{\rho}^{1-L}J_{\tum\tum}(z_\rho)$.

Meanwhile, the denominator $[q(x)/(x-z_{\rho})]_{|x=z_{\rho}}$ that appears in (\ref{dterm}), has limiting expressions that are symmetric in $z_{\rho} \to 1/z_{\rho}$. These expressions are
 \begin{eqnarray}
{[q(x)/(x-1)]}|_{x=1} \sim   \frac{\phi  (\theta +\phi ) (\theta +2 \phi )}{L^3} \label{q1} \\
{[q(x)/(x-z_{+})]}|_{x=z_{+}} \sim  {[q(x)/(x-1/z_{+})]}|_{x=1/z_{+}} \sim \frac{8 \phi^{2} }{L^{2}} \label{qzplus} \\
{[q(x)/(x-z_{-})]}|_{x=z_{-}} \sim {[q(x)/(x-1/z_{-})]}|_{x=1/z_{-}} \sim \frac{2 \phi  (\theta +\phi ) (\theta +2 \phi )}{L^3} \;.\label{qzminus}		
\end{eqnarray}
The consequence of these symmetries is that the amplitudes $b_{++}^{(0)}=b_{++}^{(1)} \equiv b_{++}$, $c_{++}^{(0)}=c_{++}^{(1)} \equiv c_{++}$,  and similarly $b_{\tum\tum}^{(0)}=b_{\tum\tum}^{(1)} \equiv b_{\tum\tum}$, $c_{\tum\tum}^{(0)}=c_{\tum\tum}^{(1)} \equiv c_{\tum\tum}$.

The remaining ingredient in the amplitudes is the leading large-$L$ behaviour of the truncated adjugate elements $\tilde{A}_{\sigma_{1}\sigma_{2},j}'(x)$ in the scaling limit. In the $++$ sector, these coincide with the expressions set out in Table~\ref{Adjugate elements in the scaling limit for constants}.  The expressions that are required in the $+-$, $+\tum$ and $\tum\tum$ sectors are provided in Tables~\ref{Adjugate elements in the scaling limit for J2}--\ref{Adjugate elements in the scaling limit for J6}. 

\begin{table}
\centering
\caption{ Adjugate elements in the scaling limit for $J_{+-}$.}
\label{Adjugate elements in the scaling limit for J2}
\begin{tabular}{r|c|c|c|c}
          		  & $x=1$ & $x=z_{+}$ & $x=1/z_{-}$ & $x=z_{-}$  \\ \hline &&&&\\[-2ex]
		  $\tilde{A}_{+-,1}$ & $\displaystyle -\frac{\theta ^2 \phi ^2 \zeta}{4 L^5}$ &  $\displaystyle  -\frac{\theta ^2 \phi ^2}{L^4}$ & $\displaystyle -\frac{\theta ^2 \phi  \zeta \left(2\lambda+\zeta \right)}{4 L^5}$  & $\displaystyle -\frac{\theta^{2}\phi \zeta(\zeta - 2\lambda)}{4L^{5}}$ \\[2ex]
$\tilde{A}_{+-,2}'$ & $\displaystyle -\frac{\theta ^2 \phi ^2 \zeta}{4 L^5}$  & $\displaystyle \frac{\theta ^3 \phi ^{3/2}}{L^{9/2}}$ & $\displaystyle \frac{\theta  \phi ^2 \zeta \left(4\lambda+\zeta + 2\eta \right)}{4 L^5}$ & $\displaystyle \frac{\theta  \phi ^2 \zeta \left(-4 \lambda+\zeta + 2\eta \right)}{4 L^5}$  \\[2ex]
$\tilde{A}_{+-,5}'$ & $\displaystyle -\frac{\theta ^2 \phi ^2 \zeta}{4 L^5}$ & $\displaystyle \frac{-\sqrt{2} \theta ^3 \phi ^{3/2}}{4 L^{9/2}}$ & $\displaystyle -\frac{\theta ^2 \phi  \zeta \left(6\lambda+2\zeta + 2\eta   \right)}{8 L^5}$  & $\displaystyle -\frac{\theta ^2 \phi  \zeta \left(-6\lambda+2\zeta + 2\eta
   \right)}{8 L^5}$   
\end{tabular}
\end{table}

\begin{table}
\centering
\caption{ Adjugate elements in the scaling limit in $J_{+\tum}$ .}
\label{Adjugate elements in the scaling limit for J5}
\begin{tabular}{r|c|c|c}
          		  & $x=1$ & $x=1/z_{-}$ & $x=z_{-}$  \\ \hline &&&\\[-2ex]
$\tilde{A}_{+\tum,1}$ & $\displaystyle -\frac{\theta  \phi ^3\zeta}{2 L^5} $ & $\displaystyle \frac{\theta ^2 \phi  \zeta \left(2\lambda+2\zeta \right)}{4 L^5}$ & $\displaystyle \frac{\theta ^2 \phi  \zeta \left(-2\lambda+2 \zeta \right)}{4 L^5}$ \\[2ex]
$\tilde{A}_{+\tum,2}'$ & $\displaystyle -\frac{\theta  \phi ^3 \zeta}{2 L^5} $ & $\displaystyle -\frac{\theta  \phi ^2 \zeta \left(6\lambda+2\zeta +2\eta
   \right)}{4 L^5}$ & $\displaystyle -\frac{\theta  \phi ^2 \zeta \left(-6\lambda+2\zeta + 2\eta
   \right)}{4 L^5}$ \\[2ex]
$\tilde{A}_{+\tum,5}'$ & $\displaystyle -\frac{\theta  \phi ^3 \zeta}{2 L^5}  $ & $\displaystyle \frac{\theta ^2 \phi  \zeta \left(4\lambda+2\zeta + \eta
   \right)}{4 L^5}$  & $\displaystyle \frac{\theta ^2 \phi  \zeta \left(-4\lambda+2\zeta + \eta
   \right)}{4 L^5}$
\end{tabular}
\end{table}

\begin{table}
\centering
\caption{Adjugate elements in the scaling limit for $J_{\tum \tum}$.}
\label{Adjugate elements in the scaling limit for J6}
\begin{tabular}{r|c|c}
          		  & $x=1$ & $x=1/z_{-}$   \\ \hline &&\\[-2ex]
$\tilde{A}'_{\tum\tum,1}$ & $\displaystyle -\frac{\phi ^4 \zeta}{L^5}$ & $\displaystyle \frac{\theta  \phi ^2 \eta (\zeta + 2 \phi )}{2 L^5}$ \\[2ex]
$\tilde{A}'_{\tum\tum,2}$ & $\displaystyle -\frac{\phi ^4 \zeta}{L^5}$  & $\displaystyle -\frac{\phi ^3 \zeta\left(2\lambda+\zeta \right)}{L^5}$ \\[2ex]
$\tilde{A}'_{\tum\tum,5}$ & $\displaystyle -\frac{\phi ^4 \zeta}{L^5}$ & $\displaystyle \frac{\theta  \phi ^2 \zeta \left(2\lambda+2\zeta
   \right)}{2 L^5}$
\end{tabular}
\end{table}

\subsection{Explicit expressions for the amplitudes in the scaling limit}
\label{section: scal limit results}

Putting this all together, we obtain explicit expressions for the amplitudes that appear in the scaling limit of the stationary probability distribution, Eqs.~(\ref{++ scaling limit})--(\ref{tum tum scaling limit}). The amplitudes that remain finite in the $L\to\infty$ limit are
\begin{eqnarray}
a_{++}  & =  \frac{\zeta  \theta ^2 P_{+\tum}(1) \left(e^{\lambda } (\eta +\lambda )-\eta +\lambda
   \right)}{4 \eta  \ell L \left(\zeta  \left(e^{\lambda }-1\right)+2 \left(e^{\lambda
   }+1\right) \lambda \right)} \\
b_{++} & =  \frac{\theta ^2 P_{+\tum}(1) }{4 L \phi }\\
c_{++}  & = \frac{\zeta  \theta ^3 e^{\lambda } P_{+\tum}(1)}{8 \ell L \phi  \left(e^{\lambda } (\eta
   +\lambda )+\eta -\lambda \right)} \\
a_{+-} & = \frac{\zeta  \theta ^2 P_{+\tum}(1)\left(e^{\lambda } (\eta +\lambda )-\eta +\lambda
   \right)}{4 \eta  \ell L \left(\zeta  \left(e^{\lambda }-1\right)+2 \left(e^{\lambda
   }+1\right) \lambda \right)} \\
c_{+-}^{(0)} & = \frac{\zeta  \theta ^3 e^{\lambda } P_{+\tum}(1)}{8 \ell L \left(e^{\lambda } \left(2
   \theta ^2+3 \theta  (\lambda +2 \phi )+4 \phi  (\lambda +\phi )\right)+\theta  \lambda
   \right)} \\
c_{+-}^{(1)} & =  -\frac{\zeta  \theta ^2 e^{\lambda } P_{+\tum}(1) (\eta +\lambda )}{4 \eta  \ell L
   \left(\zeta  \left(e^{\lambda }-1\right)+2 \left(e^{\lambda }+1\right) \lambda
   \right)}\\
a_{+\tum} & = \frac{\zeta  \theta  P_{+\tum}(1) \phi  \left(e^{\lambda } (\eta +\lambda )-\eta
   +\lambda \right)}{2 \eta  \ell L \left(\zeta  \left(e^{\lambda }-1\right)+2
   \left(e^{\lambda }+1\right) \lambda \right)} \\
c_{+\tum}^{(0)} & = \frac{\zeta  \theta ^2 e^{\lambda }  P_{+\tum}(1) (2 \lambda -\eta )}{4 \eta  \ell L
   \left(\zeta  \left(e^{\lambda }-1\right)+2 \left(e^{\lambda }+1\right) \lambda
   \right)} \\
c_{+\tum}^{(1)} & =  \frac{\zeta  \theta ^2 e^{\lambda } P_{+\tum}(1) (\eta +2 \lambda )}{4 \eta  l L
   \left(\zeta  \left(e^{\lambda }-1\right)+2 \left(e^{\lambda }+1\right) \lambda
   \right)}\\
a_{\tum \tum} & = \frac{\zeta  P_{+\tum}(1) \phi ^2 \left(e^{\lambda } (\eta +\lambda )-\eta +\lambda
   \right)}{\eta  \ell L \left(\zeta  \left(e^{\lambda }-1\right)+2 \left(e^{\lambda
   }+1\right) \lambda \right)} \\
c_{\tum \tum} & = \frac{\zeta  \theta  e^{\lambda } P_{+\tum}(1) \phi }{2 \ell L \left(e^{\lambda } (\eta
   +\lambda )+\eta -\lambda \right)} \;. 
\end{eqnarray}
All other amplitudes are zero.

The $w$ coefficients are more straightforward to obtain, since the Kronecker delta symbols $\delta_{n,1}$ and $\delta_{n,L-1}$ in (\ref{general P}) turn into Dirac delta functions $\delta(y)$ and $\delta(\ell-y)$, respectively, with their amplitudes unchanged. These amplitudes, $w_{\sigma_1\sigma_2}^{(0)}$ and $w_{\sigma_1\sigma_2}^{(1)}$ are found to be
\begin{eqnarray}
w_{+-}^{(0)} &= \frac{1}{2}(\alpha+2)\beta P_{+-}(1) + \beta^{2}P_{+-}(1) + \beta^{2}P_{+\tum}(1)/2 \\ &\sim \frac{P_{+-} (1)\theta}{L} = \frac{\theta ^2 P_{+\tum}(1) \left(e^{\lambda } (\zeta +\lambda )-\zeta +\lambda
   \right)}{L \phi  \left(\zeta  \left(e^{\lambda }-1\right)+2 \left(e^{\lambda
   }+1\right) \lambda \right)} \\
w_{+\tum}^{(0)} & = \alpha\beta P_{+-}(1) + \alpha \beta P_{+\tum}(1) + \beta P_{+\tum}(1) \sim \frac{ P_{+\tum}(0) \theta}{L} \\
w_{\tum \tum}^{(0)} & = \alpha^{2} P_{+-}(1) +(\alpha + \alpha^{2}) P_{+\tum}(1)  \sim \frac{P_{+\tum}(1) \phi}{L}\\
w_{\tum \tum}^{(1)} & = w_{\tum \tum}^{(0)} \;.
\end{eqnarray}
Again the other $w$ amplitudes are all zero.

Note that although all the amplitudes have the superficial appearance of a $1/L$ decay, this is in fact cancelled by the remaining constant, $P_{+\tum}(1)$, which scales as $L$ (as will be determined below by normalising the distribution). There is now one final remaining constant, $P_{+\tum}(1)$, which is fixed by normalisation.

\subsection{Normalisation}
\label{subsection: normalisation}

Rather than impose normalisation on the whole probability distribution, it is sufficient (and more straightforward) to impose it on a single velocity sector in order to determine $P_{+\tum}(1)$.  The relative weight of each sector can be calculated straightforwardly because {\em transitions between sectors occur at rates that are  decoupled from the hopping dynamics} i.e.~the
transitions between sectors are independent of the particle separation $n$. Moreover, each particle enters a velocity state independently of the other. Consequently, if we define the marginal probability distribution
\begin{equation}
P_{\sigma_1} = \sum_{\sigma_2} \int_0^{\ell} {\rm d} y\, P_{\sigma_1\sigma_2}(y)
\end{equation}
then  we have for the probability of being in the velocity sector $\sigma_1\sigma_2$ that
\begin{equation}
P_{\sigma_1\sigma_2} = \int_0^{\ell} {\rm d} y\,  P_{\sigma_1\sigma_2}(y) = P_{\sigma_1} P_{\sigma_2} \;.
\end{equation}

The master equation for the single particle velocity distribution reads
\begin{eqnarray}
\frac{\partial P_{+}}{\partial t} &= -\alpha P_{+} + \frac{\beta}{2} P_{\tum} \\
\frac{\partial P_{-}}{\partial t} &= -\alpha P_{-} + \frac{\beta}{2} P_{\tum} \\
\frac{\partial P_{\tum}}{\partial t} &= \alpha [P_{+} + P_{-} ] - \beta P_{\tum} \;.
\end{eqnarray}
In the steady state, we have $P_{+} = P_{-}$ by symmetry and consequently
\begin{equation}
P_{\tum} = \frac{2\alpha}{\beta} P_{+} \;.
\end{equation}
Using this result and the fact that $P_++P_\tum+P_-=1$, we find
\begin{eqnarray}
P_{+} &= P_{-} = \frac{1}{2(1+\alpha/\beta)} \;.
\end{eqnarray}
Insisting now that $\int_0^{\ell} {\rm d}\ell P_{++}(y) = P_{+}^2$, we find that
\begin{equation}
\fl P_{+\tum}(1) =L\left[(\theta +\phi )^2 \left(\zeta  \left(\frac{\sqrt{2} \theta  \left(e^{\lambda
   }-1\right)}{\phi  \sqrt{\zeta  \eta } \left(\eta +e^{\lambda } (\theta +\lambda +\phi
   )-\lambda \right)}+\frac{e^{\lambda } (\eta +\lambda )-\theta +\lambda -\phi }{\eta 
   \left(\zeta  \left(e^{\lambda }-1\right)+2 \left(e^{\lambda }+1\right) \lambda
   \right)}\right)+\frac{2}{\phi }\right)\right]^{-1},
   \end{equation}
which completes our derivation of eqs (\ref{++ scaling limit})--(\ref{tum tum scaling limit}).

\subsection{Plots of the scaling limit distribution}

\begin{figure}[h!]
  \centering
  \includegraphics[width=0.95\linewidth]{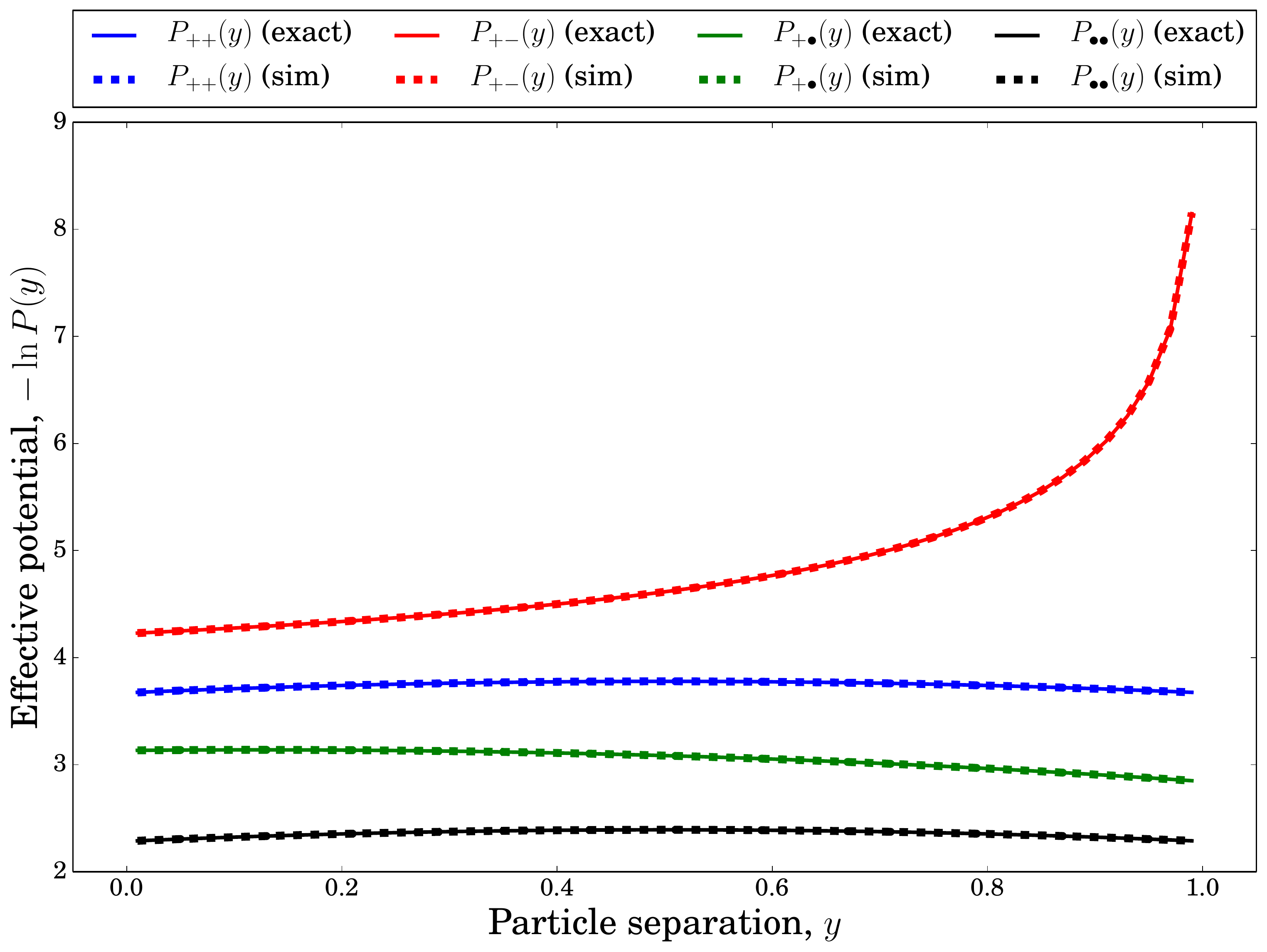}
  \caption{Comparison of exact analytic results (solid lines) with simulation results (dotted lines) for scaling limit. Model with $\phi = \theta = 1$ and $\ell = 1$. } 
    \label{fig:scal1}
  \end{figure}

\begin{figure}[h!]
  \centering
  \includegraphics[width=0.95\linewidth]{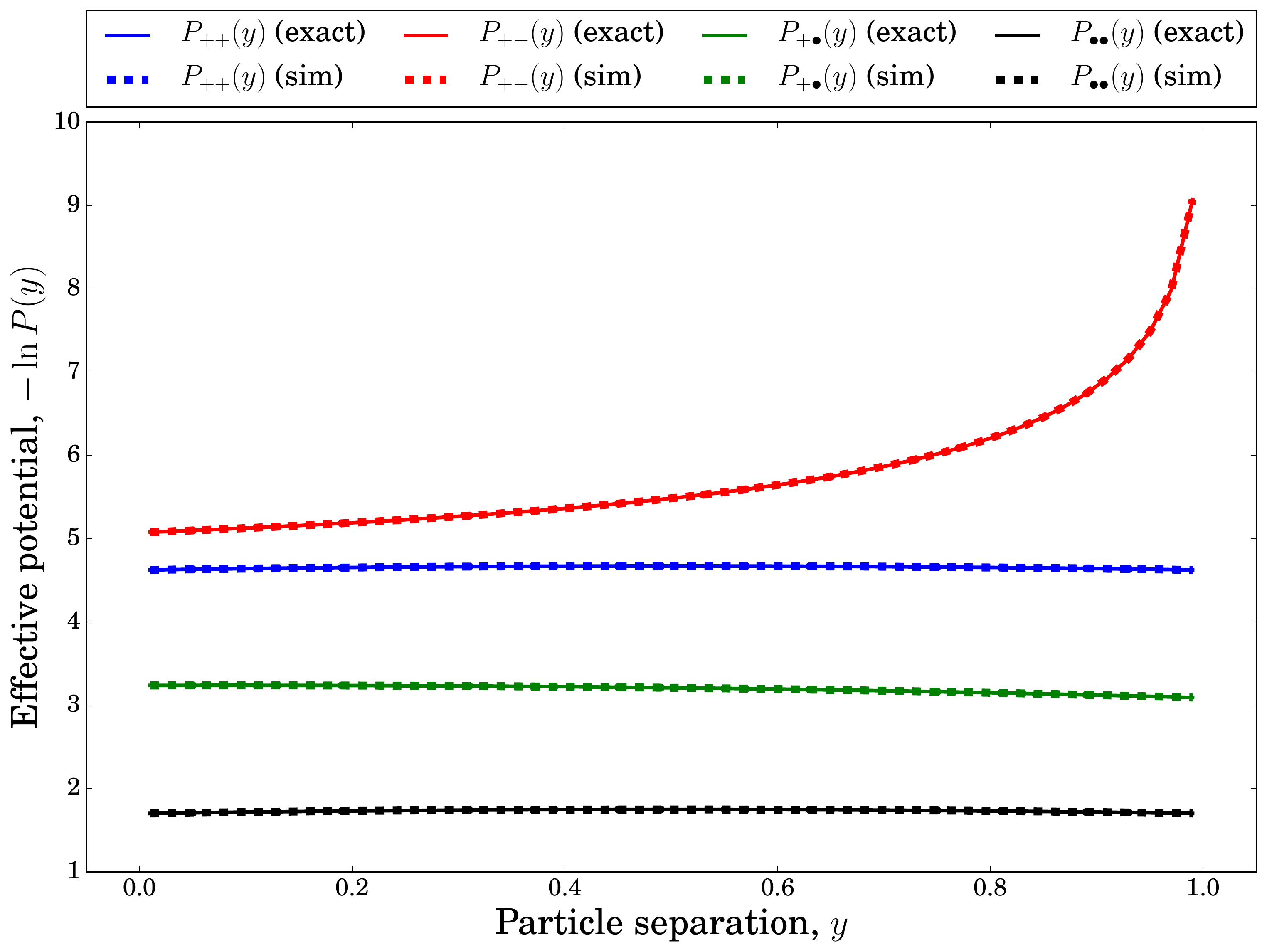}
  \caption{Comparison of exact analytic results (solid lines) with simulation results (dotted lines) for scaling limit. Model with $\phi = 1.1$, $\theta = 0.51$ and $\ell = 1$.}
  \label{fig:scal2}
  \end{figure}

We can directly simulate the scaling limit by having particles move ballistically at speed $v$ and undergoing tumbling and untumbling events at times drawn from an exponential distribution with means $1/\alpha$ and $1/\beta$ respectively. In Figures~\ref{fig:scal1} and \ref{fig:scal2} we compare the distributions (in the form of effective potentials) obtained from this simulation with our analytical calculation. Once again, we find complete agreement.  As discussed in subsection~\ref{subsection: summary of results}, and as seen explicitly above, one of the two exponential decays collapses to a delta function in this limit. Nevertheless, the second lengthscale, which is induced by the finite tumbling time, remains physically relevant in the scaling limit.


As a further check, we may consider the limit where the exit rate from tumbling $\beta \rightarrow \infty$. In this limit tumbling is instantaneous, and we recover the probability distribution in the scaling limit of the model studied in \cite{us} 
\begin{eqnarray}
P_{++}(y) =& \frac{\phi}{4\ell(4+\phi)} + \frac{ \delta (y) + \delta (\ell - y)}{2(4+ \phi)} = \frac{\tilde\alpha + 2v [ \delta(y) + \delta(\ell-y)]}{4(4v+\tilde\alpha\ell)} \label{++ beta inf} \\
P_{+-}(y) =& \frac{\phi}{4\ell (4+\phi)} + \frac{\delta (y)}{(4+\phi)} = \frac{\tilde\alpha + 4v \delta(y)}{4(4v+\tilde\alpha\ell)} \\
P_{+\tum}(y) =& P_{\tum \tum}(y)= 0.
\end{eqnarray}
Moreover, it is instructive to note exactly how this limit is recovered. 

As expected, all contributions from states with a tumbling particle vanish in this limit. There are no contributions from $c_{+-}^{(0)}e^{-y/\xi}$ and $c_{+-}^{(1)}e^{-(\ell-y)/\xi}$ as the exponentials vanish. Therefore the only terms that contribute from $(+-)$ are
\begin{eqnarray}
a_{+-} \sim & \frac{\phi}{4\ell (4+\phi)} \;\; \textrm{and,} \\
w_{+-}^{(0)} \sim & \frac{1}{4 (4+\phi)} \;\cdot
\end{eqnarray}
However, all of the terms in $(++)$ (and, equivalently, its symmetric counterpart $(--)$) do contribute to the probability in this limit. Specifically, the constant
\begin{equation}
a_{++} \sim \frac{\phi}{4\ell (4+\phi)},
\end{equation}
and 
\begin{eqnarray}
b_{++}[ \delta(y) + \delta(\ell -y)] \sim & \frac{\sqrt{2}}{4 (4+\phi)} [ \delta(y) + \delta(\ell -y)] \;\;\; \textrm{and} \\
c_{++}[e^{- y/\xi} + e^{-(\ell-y)/\xi} ] \sim & \left( \frac{1}{2(4+\phi)} - \frac{\sqrt{2}}{4(4+\phi)} \right)  [ \delta(y) + \delta(\ell -y)]. 
\end{eqnarray} 
Thus we see that not all of the probability in the delta functions in (\ref{++ beta inf}) comes from the delta-function term $b_{++}$, but that there is also a contribution from the originally finite exponential piece multiplying $c_{++}$. In other words, when the tumbling time is short (but not zero), very small inter-particle separations are generated with a high probability as a consequence of the short distance moved by a particle following a collision while the other one is tumbling.  The effect of this is seen in simulations: when $\beta$ is set very large but not strictly infinite there is a significant fraction of the probability for configurations at very marginal but non-zero separations. Only when $\beta$ is set strictly infinite does this probability moves into the delta-function terms.

\section{Conclusion}
\label{section: conclusion}

In this work, we have studied a one-dimensional lattice model of two run-and-tumble particles that tumble for  non-zero, random   amounts of time and interact under mutual exclusion. Using a generating function approach, we have exactly solved the stationary distribution of the particle positions and velocities. Our results, visualised in Figs.~\ref{lattice model prob} and \ref{fig:scal1} in the form of effective potentials, show that effective attractions emerge. Physically, we can understand this as being due to particle collisions. On colliding, the particles jam until one of them tumbles and then moves away, which causes probability to accumulate in configurations where particles oppose each other on neighbouring sites. Mathematically, this is represented by the delta symbol contributions in Eq.~(\ref{P intro}). We also find this type of delta symbol contribution where particles are on adjacent sites and are both tumbling. This is due to the the high probability of entering this configuration from jamming collisions.

This jamming of the particles is in turn responsible for the rest of the structure of the probability distribution. We found this to be characterised by two lengthscales
($[ \ln z_+ ]^{-1}$ and  $[ \ln z_-]^{-1}$ where $z_+$ and $z_-$ appear in 
expression (\ref{P intro}) for the stationary distribution). The first of these
($[ \ln z_+ ]^{-1}$) can be attributed to fluctuations in the separation between the two particles due to their stochastic hopping. The contribution to the probability from this broadening after unjamming decays exponentially as the separation between the particles increases. We can ascribe this lengthscale to the stochastic hopping because it vanishes in the scaling limit in which the motion becomes ballistic. Moreover, this lengthscale is also present in the limit where tumbling is instantaneous \cite{us}.

The second lengthscale ($[ \ln z_- ]^{-1}$) appears in those cases where the tumbling process has a finite mean time. In particular, this generates configurations in which one particle is tumbling whilst the other particle moves. The typical distance travelled by a particle in such configurations remains finite in the scaling limit, and consequently the second lengthscale is also finite in this limit. This lengthscale depends on a combination of both the tumbling entry and tumbling exit rates, as together they determine how far the moving particle may separate itself from the stationary particle. It furthermore appears in all the velocity sectors.

Together these results demonstrate the rich structure that non-equilibrium stationary states may exhibit, even in relatively simple systems where detailed balance is broken.  In this work, we built on our earlier study of a similar model in which particles tumbled instantaneously (persistent random walkers) \cite{us}, motivated by experimental observations that the tumbling time is a random variable that is reasonably well described by an exponential distribution with a finite mean \cite{Berg2004}. 
 We have seen that this additional feature of the microscopic dynamics
has led to  the appearance of a new lengthscale  which survives in the scaling limit. 
In principle, changes in the microscopic dynamics could lead to additional structure 
  entering  the stationary distribution in a variety of ways, as we now discuss.

To understand other possible structures for stationary states, it is worth delving a little more deeply into the mathematical structure of the solution we have presented. A crucial step is the inversion of the matrix $A$ (\ref{matG}) that relates the generating functions in each velocity sector to one another.  The elements of this matrix contain terms proportional to the generating function variable $x$ or to its reciprocal, $1/x$.  This is due to particles hopping one site at a time (if they could hop two sites, one would obtain $x^2$ and $1/x^2$, and so on). The consequence of this is that the elements of the inverse matrix $A^{-1}$ can be written as the ratio of two polynomials, each related to the determinant of $A$ or one of its submatrices. If the numerator polynomial is of lower degree than the denominator polynomial, the generating function has simple poles which, on inversion, translate to exponential decays in the stationary probability distribution. On the other hand, if the numerator polynomial has the same or higher degree than the denominator polynomial, there are additional (`anomalous') contributions corresponding to  particle separations that are determined by the difference in the degree of the two polynomials. It is not obvious that the addition of an `internal' process to the particle dynamics (stochastic switching between a running and a tumbling state without changing its position) should be of the type that generates an extra lengthscale rather than anomalous contributions to the probability distribution. It would be interesting to understand more deeply the structure of the $A$ matrix and thereby what physical processes tend to create effective inter-particle interactions of different types. 

More broadly there is scope to incorporate additional features of real bacterial dynamics into the model. The most obvious directions for further study would be in increasing the number of particles and the dimensionality of the system. In the former case, it would be interesting to determine whether an effective interaction between three (or more) particles can be decomposed into two-body interactions. In the latter, one would like to know, for example, whether the short range attraction that is mediated by jamming survives. The greatest insights are probably to be gained if both generalisations are combined; 
however solving a model of this complexity remains a theoretical challenge.

\vspace{1cm}

\bibliography{MyCollection}
\bibliographystyle{unsrt}

\end{document}